\documentclass[12pt]{article}
\usepackage{amssymb}
\usepackage{epsfig}
\usepackage{amsmath}
\usepackage[utf8x]{inputenc}
\usepackage{amssymb}
\usepackage{slashed}
\usepackage{marvosym}
\usepackage{graphicx}
\usepackage{color}
\voffset= - 0.3in
\hoffset= - 0.7in         

\catcode`\@=11
\def\marginnote#1{}

\newcount\hour
\newcount\minute
\newtoks\amorpm
\hour=\time\divide\hour by60
\minute=\time{\multiply\hour by60 \global\advance\minute by-\hour}
\edef\standardtime{{\ifnum\hour<12 \global\amorpm={am}%
        \else\global\amorpm={pm}\advance\hour by-12 \fi
        \ifnum\hour=0 \hour=12 \fi
        \number\hour:\ifnum\minute<10 0\fi\number\minute\the\amorpm}}
\edef\militarytime{\number\hour:\ifnum\minute<10 0\fi\number\minute}

\def\draftlabel#1{{\@bsphack\if@filesw {\let\thepage\relax
   \xdef\@gtempa{\write\@auxout{\string
      \newlabel{#1}{{\@currentlabel}{\thepage}}}}}\@gtempa
   \if@nobreak \ifvmode\nobreak\fi\fi\fi\@esphack}
        \gdef\@eqnlabel{#1}}
\def\@eqnlabel{}
\def\@vacuum{}
\def\draftmarginnote#1{\marginpar{\raggedright\scriptsize\tt#1}}

\def\draft{\oddsidemargin -.5truein
        \def\@oddfoot{\sl preliminary draft \hfil
        \rm\thepage\hfil\sl\today\quad\militarytime}
        \let\@evenfoot\@oddfoot \overfullrule 3pt
        \let\label=\draftlabel
        \let\marginnote=\draftmarginnote
   \def\@eqnnum{(\theequation)\rlap{\kern\marginparsep\tt\@eqnlabel}%
\global\let\@eqnlabel\@vacuum}  }

\def\appname{Appendix}
\newcounter{app}
\def\theapp{\Alph{app}}
\def\app{\par
   \addvspace{4ex}
   \@afterindentfalse
  \secdef\@app\@dapp}
\def\@app[#1]#2{\ifnum \c@secnumdepth >\m@ne
        \refstepcounter{app}
        \addcontentsline{toc}{app}{\theapp
        \hspace{1em}#1}\else
      \addcontentsline{toc}{app}{ #1}\fi
   {\parindent \z@ \raggedright
    \Large \bf \appname~\theapp .
   \Large  \bf 
    #2}\nobreak
   \vskip 4ex   \noindent
\setcounter{equation}{0}
\def\theequation{\Alph{app}.\arabic{equation}}}
\def\@dapp#1{%
{\parindent \z@ \raggedright  \bf #1}\par\nobreak}
\def\l@app#1#2{\addpenalty{\@secpenalty}%
   \addvspace{1em plus\p@}%
   \begingroup
   \@tempdima 3em
     \parindent \z@ \rightskip \@pnumwidth
     \parfillskip -\@pnumwidth
     { \bf
     \leavevmode
     #1\hfil \hbox to\@pnumwidth{\hss #2}}\par
     \nobreak
   \endgroup}

\makeatletter
\newdimen\normalarrayskip            
\newdimen\minarrayskip               
\normalarrayskip\baselineskip
\minarrayskip\jot
\newif\ifold             \oldtrue            \def\new{\oldfalse}
\def\arraymode{\ifold\relax\else\displaystyle\fi}
\def\eqnumphantom{\phantom{(\theequation)}} 
\def\@arrayskip{\ifold\baselineskip\z@\lineskip\z@
     \else
     \baselineskip\minarrayskip\lineskip1\baselineskip\fi}

\def\@arrayclassz{\ifcase \@lastchclass \@acolampacol \or
\@ampacol \or \or \or \@addamp \or
   \@acolampacol \or \@firstampfalse \@acol \fi
\edef\@preamble{\@preamble
  \ifcase \@chnum
     \hfil$\relax\arraymode\@sharp$\hfil
     \or $\relax\arraymode\@sharp$\hfil
     \or \hfil$\relax\arraymode\@sharp$\fi}}

\def\@array[#1]#2{\setbox\@arstrutbox=\hbox{\vrule
     height\arraystretch \ht\strutbox
     depth\arraystretch \dp\strutbox
width\z@}\@mkpream{#2}\edef\@preamble{\halign \noexpand\@halignto
\bgroup \tabskip\z@ \@arstrut \@preamble \tabskip\z@ \cr}%
\let\@startpbox\@@startpbox \let\@endpbox\@@endpbox
  \if #1t\vtop \else \if#1b\vbox \else \vcenter \fi\fi
  \bgroup \let\par\relax
  \let\@sharp##\let\protect\relax
  \@arrayskip\@preamble}
\def\eqnarray{\stepcounter{equation}%
              \let\@currentlabel=\theequation
              \global\@eqnswtrue
              \global\@eqcnt\z@
              \tabskip\@centering              
              \let\\=\@eqncr
              $$%
            \halign to \displaywidth  \bgroup
             \eqnumphantom \@eqnsel
      \hskip\@centering                               
    $\displaystyle  \tabskip\z@ {##}$%
    &\global\@eqcnt\@ne \hskip 2\arraycolsep
         $ \displaystyle  \arraymode{##}$\hfil
    &\global\@eqcnt\tw@ \hskip 2\arraycolsep
         $\displaystyle\tabskip\z@{##}$\hfil
         \tabskip\@centering
    &{##}\tabskip\z@\cr}
\makeatother

\newfont{\hr}{msbm10}
\newfont{\ams}{msam10}

\font\numbers=cmss12
\font\upright=cmu10 scaled\magstep1
\def\stroke{\vrule height8pt width0.4pt depth-0.1pt}
\def\topfleck{\vrule height8pt width0.5pt depth-5.9pt}
\def\botfleck{\vrule height2pt width0.5pt depth0.1pt}
\def\Zmath{\vcenter{\hbox{\numbers\rlap{\rlap{Z}\kern 0.8pt\topfleck}\kern
2.2pt
                   \rlap Z\kern 6pt\botfleck\kern 1pt}}}
\def\Qmath{\vcenter{\hbox{\upright\rlap{\rlap{Q}\kern
                   3.8pt\stroke}\phantom{Q}}}}
\def\Nmath{\vcenter{\hbox{\upright\rlap{I}\kern 1.7pt N}}}
\def\Cmath{\vcenter{\hbox{\upright\rlap{\rlap{C}\kern
                   3.8pt\stroke}\phantom{C}}}}
\def\Rmath{\vcenter{\hbox{\upright\rlap{I}\kern 1.7pt R}}}
\def\Z{\ifmmode\Zmath\else$\Zmath$\fi}
\def\Q{\ifmmode\Qmath\else$\Qmath$\fi}
\def\N{\ifmmode\Nmath\else$\Nmath$\fi}
\def\C{\ifmmode\Cmath\else$\Cmath$\fi}
\def\R{\ifmmode\Rmath\else$\Rmath$\fi}

\def\d{\partial}

\def\bea{\begin{eqnarray}}
\def\eea{\end{eqnarray}}

\def\beq{\begin{equation}}
\def\eeq{\end{equation}}
\def\ba{\beq\new\begin{array}{c}}
\def\ea{\end{array}\eeq}
\def\be{\ba}
\def\ee{\ea}
\def\F{{\cal F}}

\def\Q{{\cal Q}}

\def\stackreb#1#2{\mathrel{\mathop{#2}\limits_{#1}}}

\def\res{{\rm res}}

\def\half{{\textstyle{1\over2}}}

\def\N2{${\cal N}=2$}
\def\4N{${\cal N}=4$}
\def\1N{${\cal N}=1$}
\def\1N*{${\cal N}=1^*$}

\def\beq{\begin{equation}}
\def\eeq{\end{equation}}
\def\ba{\beq\new\begin{array}{c}}
\def\ea{\end{array}\eeq}
\def\be{\ba}
\def\ee{\ea}
\def\theequation{\thesection.\arabic{equation}}

\newcommand{\rf}[1]{(\ref{#1})}

\newcommand{\tr}{\mathrm{tr\,}}
\newcommand{\pd}{\partial}

\newcommand{\bpm}{\begin{pmatrix}}
\newcommand{\pbm}{\end{pmatrix}}
\newcommand{\Sum}{\sum\limits}
\newcommand{\gb}{\end{gathered}}
\newcommand{\Int}{\int\limits}

\newcommand{\Oint}{\oint\limits}

\newtheorem{theorem}{Theorem}

\begin{document}


\begin{flushright}
FIAN/TD-20/13\\
ITEP/TH-47/13
\end{flushright}
\vspace{1.0 cm}

\begin{center}
\baselineskip30pt
{\bf \LARGE Residue Formulas for Prepotentials, Instanton Expansions
and Conformal Blocks}
\end{center}
\bigskip
\bigskip
\begin{center}
\baselineskip12pt
{\large P.~Gavrylenko~$^{a,b}$ and A.~Marshakov~$^{c,a}$}\\
\end{center}
\bigskip

\begin{center}
$^a${\it Department of Mathematics and Laboratory of Mathematical \\
Physics, NRU HSE, Moscow, Russia}\\
$^b${\it Bogolyubov Institute for Theoretical Physics and
Department \\ of Physics, Kyiv National University, Ukraine}\\
$^c${\it Theory Department, Lebedev Physics Institute and
Institute for \\ Theoretical and Experimental Physics, Moscow, Russia}\\

\end{center}
\bigskip\medskip

\begin{center}
{\large\bf Abstract} \vspace*{.2cm}
\end{center}

\begin{quotation}
\noindent
We study the extended prepotentials for the S-duality class of quiver gauge theories,
considering them as
quasiclassical tau-functions, depending on gauge theory condensates and bare couplings.
The residue formulas for the third derivatives of extended prepotentials are proven,
which lead to effective way of their computation, as expansion in
the weak-coupling regime. We discuss also the differential equations, following from the residue formulas, including the WDVV equations, proven to be valid for the $SU(2)$ quiver gauge theories. As a particular
example we consider the constrained conformal quiver gauge theory, corresponding to the Zamolodchikov conformal blocks by 4d/2d duality. In this
case part of the found differential equations turn into nontrivial relations for the
period matrices of hyperelliptic curves.
\end{quotation}

\setcounter{equation}0
\section{Introduction
\label{ss:intro}}

The S-duality class of the supersymmetric quiver theories \cite{gaiotN2dual} allows to
study gauge theories with matter, charged with respect to more than a single gauge group. In some
regions of their moduli space the traditional methods of quantum field theory are not
applicable, but they can be still analyzed, using geometric approach to \N2 supersymmetric gauge
theories, initiated long ago by Seiberg and Witten \cite{SW}. It is especially intriguing and
interesting, that this geometry can be independently viewed both from four-dimensional and two-dimensional
perspectives \cite{LMN,NO,BE,AGT}, allowing to apply in the latter case the dual techniques of theories
with infinite-dimensional algebras of symmetry. Direct observation of such symmetry
in four-dimensional gauge theories remains to be one of the main challenges in modern mathematical
physics.

The Seiberg-Witten (SW) prepotentials in quiver gauge theories can be naturally extended \cite{AMtau}
to incorporate the dependence of the bare ultraviolet (UV) couplings in addition to the infrared (IR) gauge theory condensates.
In this way they can be identified with more general class of the quasiclassical tau-functions \cite{KriW}, which are well-known from long ago \cite{GKMMM} to appear in the context of supersymmetric gauge theories.
Studied previously only for the higher perturbations of the UV prepotential \cite{LNS,GMMMRG,LMN,MN}, this extension becomes more generic for the quiver theories and can be studied in detail
along the lines, proposed in \cite{AMtau}.

One of the practical applications of the geometric picture and integrable equations in the gauge-theory framework is
that it allows to use the alternative methods for the computations of the prepotentials, describing
physics of light states in the IR. These methods equally work in the regions, where the theory possesses the UV non-Abelian Lagrangian description, as well as in the regions where only an effective description of light Abelian modes is possible (or even no Lagrangian is known at all \cite{gaiotN2dual}). In the first case
the weak-coupling phase in \N2 supersymmetric gauge theory is saturated by the one-loop perturbation theory and the instanton calculus \cite{SW,Nek}, while in the latter cases the traditional methods of quantum gauge theory are not fully applicable. Below we are going to use the techniques of the quasiclassical integrable systems \cite{KriW} to study the expansion
of the prepotentials of quiver gauge theories in various regions of the moduli space, to be called
as \emph{weak-coupling expansion}, since it coincides with the perturbative and instanton analysis in all known cases, though goes itself beyond the scope, where such analysis is valid.

In this paper the proposed methods will be used to study the prepotentials of S-duality class of the $SU(2)$ quiver gauge theories, and mostly with the massless matter (vanishing bare masses). Already in this case there are some subtleties with the analysis
of weakly-coupled phases with (half-) tri-fundamental multiplets (the so called sicilian quivers),
where the instanton calculus \cite{Nek} is
not directly applicable \cite{nepes}. Nevertheless, the developed methods lead directly to the weak-coupling expansion in this case as well, and this allows to hope on their
validity for the most interesting case of the higher-rank gauge quivers, where the ``sicilian problem'' arises in full. Moreover, when one of the trifundamental states becomes massless, such theories form a class of four-dimensional conformal theories with the quadratic prepotentials, when the couplings are renormalized from their bare values only by non-perturbative effects (the perturbative beta-functions vanish and the perturbative renormalization is finite).

Fortunately, for this class of quiver theories there is a well-known description on two-dimensional side,
proposed by Al.~Zamolodchikov in terms of conformal blocks for the $c=1$ Ashkin-Teller (AT) model
or scalar field on hyperelliptic Riemann surface \cite{ZamAT}. In the SW approach this Riemann surface appears
just as a particular degenerate case of the SW curve for a massless $SU(2)$ quiver, when the gauge theory condensates are constrained by certain
conservation condition. We establish here direct relation of the geometric approach with the formulation in terms of two-dimensional conformal field theory (2d CFT), which is one more nice example of 4d/2d correspondence, going - strictly speaking - even
beyond the framework of the AGT conjecture \cite{AGT}, since generic Nekrasov function is now known
in such cases \cite{nepes} (see also \cite{siqui}).

We derive a generalization of the Zamolodchikov renormalization formula \cite{ZamCB} (see also \cite{ZAGTKl,ZAGT,Pogh,Egumar,AMtau}) for this class of constrained
quiver theories, which includes the first-order differential equations for the effective couplings and their implicit solution via the Thomae formulas. Finally, we study another kind of non-linear differential equations for the extended prepotentials - the WDVV equations \cite{WDVV,Dub,MMM}, directly following from the residue formulas, and show that they hold both in generic massless and constrained Zamolodchikov's cases.

This paper is organized as follows. Sect.~\ref{ss:gen} contains the review of the SW approach, and its extension to the tau-functions of the quiver gauge theories.
We also demonstrate here, that part of the formulas can be immediately derived, using the AGT correspondence with conformal theory, and prove existence of the standard residue formula \cite{KriW} for the third derivatives of the extended quiver prepotentials.

In sect.~\ref{ss:wcexp} we propose two slightly different, but directly related methods of calculating
the weak-coupling expansions of the prepotentials, based on differential equations, arising from the residue formulas and the period integral expansions. We reproduce in this way few well-known examples,
and study in detail two quiver gauge models (including the case of sicilian quiver in
sect.~\ref{ss:quiver}) and massive
deformations of conformal gauge theories.
Sect.~\ref{ss:AlZam} contains the analysis of connection between the constrained quiver theories and exact Zamolodchikov's conformal blocks in AT model, we prove that the SW
description coincides with the 2d CFT result. Sect.~\ref{ss:eqs} contain the discussion of the nonlinear differential equations for the quiver tau-functions. We show, that the simplest relations for the period matrices of Zamolodchikov's case are equivalent to the well-known Rauch formulas, which describe here the nonperturbative renormalization of the UV couplings in conformal theories. This section also contains the proof that the prepotential solves the WDVV equations \cite{MMM} as the function of \textit{all} variables: both in the full massless theory,
and in the constrained case. Sect.~\ref{ss:concl} is devoted to brief discussion of our results. Some technical information is contained in Appendices.

\setcounter{equation}0
\section{Generalities
\label{ss:gen}}

\subsection{Integrability}

We start from the definition of the SW system \cite{SW}, assuming here the S-duality class of
quiver gauge theories \cite{gaiotN2dual} with the $\bigotimes_k SU(N_c^{(k)})$ gauge groups. We shall mostly concentrate on the superconformal models, containing fundamental, bifundamental and trifundamental matter multiplets in a combination, which gives vanishing $\beta$-functions $\beta_k=0$ for all simple
gauge group components, moreover - often with vanishing bare masses. The asymptotically free cases can be obtained from superconformal
models with dimensionless bare couplings by standard dynamical transmutation in the limit when (some
of) the bare masses become infinite and corresponding multiplets decouple.

The definition includes:

\begin{itemize}
\item $g$-parametric family of the genus $g$ curves $\Sigma$: $F(x,z;u_1,...,u_g)=0$ with the fixed
basis in $H_1(\Sigma,\mathbb{Z})$ (including marked $A$- and $B$-cycles).
\item Two meromorphic differentials $dx$ and $dz$ or the SW one-form $dS=x dz$.
\item The connection $\nabla$ on the moduli space, such that derivatives $\nabla_{\frac{\pd}{\pd u_i}}dS$ are holomorphic.
\end{itemize}

The SW equations read (see \cite{KriW} for the most general construction and \cite{SW} for the first application to supersymmetric
 gauge theory)
\be
a_i=\frac1{2\pi i}\Oint_{A_i}dS,\hspace{1cm} a^D_i=\Oint_{B_i}dS = \frac{\pd\mathcal{F}}{\pd a_i}\ \ \ \ i=1,\ldots,g
\label{SW0}
\ee
\begin{theorem}There exists locally-defined function $\mathcal{F}(a_1,...a_g)$,
which solves this system, the difference of any two solutions is $\mathbf{a}$-independent.
\end{theorem}
\textbf{Proof:} Denote $\nabla_{\frac{\pd}{\pd a_i}}=\frac{\pd}{\pd a_i}$ and compute the derivatives $$\delta_{ij}=\frac1{2\pi i}\Oint_{A_i}\frac{\pd dS}{\pd a_j}\hspace{1cm}\frac{\d a^D_i}{\d a_j}
=\Oint_{B_i}\frac{\pd dS}{\pd a_j}=\mathcal{T}_{ij}$$
We conclude from the first formula that $\frac{\pd dS}{\pd a_i}=d\omega_i$, $i=1,\ldots,g$ are canonically-normalized holomorphic one-forms, so the
second derivatives of the prepotential are the period matrix, which is symmetric due to Riemann bilinear relations (RBR)
\be
\label{RBIhol}
0 = \int_{\Sigma} d\omega_i\wedge d\omega_j = \int_{\d\Sigma_{\rm cut}} \omega_id\omega_j
= \sum_{k=1}^g \left(\oint_{A_k}d\omega_i\oint_{B_k}d\omega_j -
\oint_{A_k}d\omega_j\oint_{B_k}d\omega_i\right) =
\\
= \mathcal{T}_{ij} - \mathcal{T}_{ji}
\ee
where we have used the Stokes formula for
the integral over the boundary $\d\Sigma_{\rm cut}$ of the cut Riemann surface $\Sigma$.
Hence, we have proven that 1-form on the moduli space $\eta=\Sum_{i=1}^g \delta a_i a^D_i=\Sum_{i=1}^g \delta\left(\Oint_{A_i}dS\right)\cdot\Oint_{B_i}dS$ is closed, so locally it is the
differential of some function $\mathcal{F}$.

 \hfill $\Box$

Now let us extend and precise our definition, for the family of curves
\be
\label{famcur}
F(x,z;\mathbf{u},\mathbf{q}) =F(x,z;u_1,\ldots,u_g;q_1,\ldots,q_n) = \sum_k x^k\phi_k(z) = 0
\ee
which covers $\Sigma\to\Sigma_0$ some curve $\Sigma_0$ (which is often called UV or Gaiotto curve), whose moduli space
can be parameterized by $\mathbf{q}=\{q_1,...,q_n\}$, while $\mathbf{u}=\{u_1,\ldots,u_g\}$
are moduli
of the cover~\footnote{In the most simple, but still important case the cover is hyperelliptic, defined by the
quadratic equation $x^2=\phi_2(z)$, with the poles at $n$ marked points. In this case the number of vacuum condensates (the
dimension of the space of such differentials) equals to the dimension of Teichm\"uller space of $\Sigma_0$:
$l(2K+p_1+...+p_n)=4g_0-4+n-g_0+1=3g_0-3+n=\mathrm{dim\,}\mathrm{Teich}_{g_0,n}$, i.e. the number of coupling constants equals to the
number of vacuum condensates (each $SU(2)$ gauge group comes with the coupling constant
and the vacuum condensate), and the whole set of parameters can be identified with $T^\ast(\mathrm{Teich}_{g_0,n})$. The genus of the cover $\Sigma$ comes from the Riemann-Hurwiz formula: $g=2g_0-1+
\frac12\Sum(n_i-1)=2g_0-1+2g_0-2+n=g_0+(3g_0-3+n)$. Unfortunately, only partially such analysis can be applied to the case of higher-rank gauge quivers, see \cite{AMtau}. However, we shall also consider below the Zamolodchikov or constrained case, where the number of gauge theory condensates is constrained by certain conservation conditions (or vanishing of the masses of some light physical states), but the
number of UV couplings remains intact, then the reduced genus $\tilde{g}<n$.}. The curve $\Sigma$ is endowed with two meromorphic differentials \cite{KriW}: $dz$, which can
be projected to the UV curve $\Sigma_0$ and $dx$, or generating one-form $dS=xdz$.
The connection $\nabla=\nabla^z$ can be now defined via projection
onto the $z$-plain:
$$\nabla_{\frac{\pd}{\pd u_i}}f(z,x(z, \mathbf{u}))dz=\frac{\pd f(z,x(z, \mathbf{u})}{\pd u_i}dz$$
For the rational $\Sigma_0$ with $g_0=0$ parameters $\{q_i\}$ are the co-ordinates
of the following points in $z$-plane (both options are possible):
\begin{itemize}
\item The branch-points of the cover, where $x\mathop{=}\limits_{z\to q_i}\Sum_{l=1}^{k-1}C_l(z,{\bf q},\vec u)(z-q_i)^{-\frac lk}$ (massless case).
\item The set of the regular points on the cover $\Sigma$, where $xdz$ has the simple poles with fixed residues (massive case).
\end{itemize}

Consider now the following system of equations:
\begin{equation}
\label{SW1}
\frac{\pd\mathcal{F}}{\pd q_i}=\frac12\Sum_{p^{-1}(q_i)}\mathrm{Res}\,\frac{\left(dS\right)^2}{dz},
\ \ \ \ i=1,\ldots,n
\end{equation}
Here we should compute the number of points with their multiplicities. Then there is the non-trivial

\begin{theorem}
The systems (\ref{SW0}) and (\ref{SW1}) are consistent. They define $\mathcal{F}(\mathbf{a},\mathbf{q})$, which is defined
up to some constant, whereas (\ref{SW0}) defines $\mathcal{F}(\mathbf{a},\mathbf{q})$ up to some $q_i$-dependent function.
\end{theorem}
\textbf{Proof:}
In the vicinity of branching points one has
\be
dS^{(1)}\ \stackreb{z\to q_i}{=}\ \Sum_{l=1}^{k-1}\frac{C_l(z,{\bf q},{\bf a})dz}{(z-q_i)^{\frac lk}}\\
dS^{(2)}\ \stackreb{z\to q_i}{=}\ \frac{m_i dz}{z-q_i}+C_0(z,{\bf q},{\bf a})dz
\ee
where two different possibilities are marked by $^{(1)}$ or $^{(2)}$ respectively.
Here $C_l(z,{\bf q},{\bf a})=C_l+c_{1l}(z-q_i)+c_{2l}(z-q_i)^2+\ldots$ are analytic at $z\to q_i$,
therefore
\be
\frac{\pd dS^{(1)}}{\pd q_i}\mathop{=}\limits_{z\to q_i}\Sum_{l=1}^{k-1}\frac lk
\frac{C_l dz}{(z-q_i)^{\frac lk+1}}+hol.\mathop{=}\limits_{z\to q_i}
-d\Sum_{l=1}^k\frac{C_l dz}{(z-q_i)^\frac lk}+hol.\\
\frac{\pd dS^{(2)}}{\pd q_i}\mathop{=}\limits_{z\to q_i}\frac{m_i dz}{(z-q_i)^2}+hol.\mathop{=}\limits_{z\to q_i}-d\frac{m_i}{z-q_i}+hol.
\ee
since the residues are fixed, or $\frac{\pd m_i}{\pd q_i}=0$.
Denote $\frac{\pd dS}{\pd q_i}=d\Omega_i$, then in
both cases
\be
\frac{dS}{dz}\mathop{=}\limits_{z\to q_i}-\Omega_i+hol.\label{dSdz}
\ee
where $\Omega_i(P)=\Int_{P_0}^{P}d\Omega_i$ is corresponding Abelian integral.
For the mixed derivatives one gets from \rf{SW1}
\be
\frac{\pd\mathcal{F}}{\pd a_j \pd q_i}=\Sum_{p^{-1}(q_i)}\mathrm{Res}\,\frac{dS}{dz}\frac{\pd dS}{\pd a_j}=
\Sum_{p^{-1}(q_i)}\mathrm{Res}\frac{dS}{dz}d\omega_j=-
\Sum\mathrm{Res}(\Omega_id\omega_j)
\ee
where the replacement $\frac{dS}{dz}\mathop{\approx}\limits_{z\to q_i}-\Omega_i$ is allowed, because $d\omega_j$ is holomorphic.
On the other hand from \rf{SW0}
\be
\frac{\pd^2\mathcal{F}}{\pd q_i\pd a_j}=\Oint_{B_j}\frac{\pd dS}{\pd q_i}=\Oint_{B_j}d\Omega_i=
\frac1{2\pi i}\Sum_{l=1}^g\left(\Oint_{A_l}d\omega_j\Oint_{B_l}d\Omega_i-
\Oint_{B_l}d\omega_j\Oint_{A_l}d\Omega_i \right) =
\\
= -\frac1{2\pi i}
\oint_{\d\Sigma_{\rm cut}}\Omega_id\omega_j =-
\Sum\mathrm{Res}\left(\Omega_id\omega_j\right)
\ee
where we have used $0=\frac{\pd a_k}{\pd q_i}=\Oint_{A_k}d\Omega_j$,
and we sum over all poles of $\Omega_i$. So we have proven that $\frac{\pd^2\mathcal{F}}{\pd a_j\pd q_i}=
\frac{\pd^2\mathcal{F}}{\pd q_i\pd a_j}$, which means the consistency of equations.

Now consider the second set of the mixed derivatives
\be
\label{2ndq}
\frac{\pd^2\mathcal{F}}{\pd q_j\pd q_i}=\Sum_{p^{-1}(q_i)}\mathrm{Res}\,\frac{dS}{dz}d\Omega_j=
-\Sum_{p^{-1}(q_i)}\mathrm{Res}(\Omega_i d\Omega_j)
\ee
which gives for the difference
\be
\frac{\pd^2\mathcal{F}}{\pd q_j\pd q_i}-\frac{\pd^2\mathcal{F}}{\pd q_i\pd q_j}=-
\Sum_{p^{-1}(q_i)}\mathrm{Res}
(\Omega_id\Omega_j)+\Sum_{p^{-1}(q_j)}\mathrm{Res}(\Omega_jd\Omega_i)=\\=-\Sum_{p^{-1}(q_i)}\mathrm{Res}
\,d(\Omega_i\Omega_j)+\Sum\mathrm{Res}\left( \Omega_jd\Omega_i\right)
\ee
Here the first term is zero due to the trivial reason, and second is zero due to
\be
\Sum\mathrm{Res}\left( \Omega_jd\Omega_i\right) =
\frac1{2\pi i}\Sum_{l=1}^g\left(\Oint_{A_l}d\Omega_j\Oint_{B_l}d\Omega_i-
\Oint_{B_l}d\Omega_j\Oint_{A_l}d\Omega_i \right) = 0
\ee
Hence, all mixed second derivatives are equal due to the RBR.\hfill $\Box$

\subsection{Residue formula
\label{ss:res}}

The third derivatives of quasiclassical tau-functions should be generally presented by the elegant
residue formulas \cite{KriW}. To prove it for our case
we unify all variables into a single set $\{ X_I\}=\{ a_i\}\cup\{ q_k\}$, and the same with
the differentials: $\{ d\varpi_I\}=\{ d\omega_i\}\cup\{ d\Omega_k\}$. 

\begin{theorem}There is a set of residue formulas for the third derivatives of the generalized
prepotential defined by (\ref{SW0}) and (\ref{SW1})
\be
\frac{\pd^3\mathcal{F}}{\pd X_I\pd X_J\pd X_K}=\Sum_{dx=0}{\rm Res}\frac{d\varpi_I d\varpi_J d\varpi_K}
{dx dz}
\label{residue}
\ee
\end{theorem}

\textbf{Proof:} Let us consider the most subtle case of the third $\mathbf{q}$-derivatives
\be
\label{residue0}
\frac{\pd^3\mathcal{F}}{\pd q_i\pd q_j\pd q_k}=\Sum_{dx=0}\mathrm{Res}\,\frac{d\Omega_i d\Omega_j d\Omega_k}{dx dz}
\ee
Formulas for the third $\mathbf{a}$-derivatives (see, e.g. \cite{ploh}) and mixed derivatives are proven just in the same way.

Start with
\be
\frac{\pd^2\mathcal{F}}{\pd q_j\pd q_i}=\Sum_{p^{-1}(q_i)}\mathrm{Res}\,\frac{dS}{dz}d\Omega_j=
\Sum_{p^{-1}(q_i)}\mathrm{Res}\, xd\Omega_j
\ee
To take extra $q$-derivative it is more convenient to use connection $\nabla^x$, which is defined via the projection onto the $x$-plane:
\be
\nabla^x_{\d\over\d q_i} d\Omega_j = {\d\over\d q_i}d\Omega_j(x,z(x, \mathbf{u}))=
\left.{\d\over\d q_i}d\Omega_j\right|_x
\ee
so that
\be
\label{respre}
\frac{\pd^3\mathcal{F}}{\pd q_j\pd q_i\pd q_k}=
\Sum_{p^{-1}(q_i)}\mathrm{Res}\,\left. x\frac{\pd d\Omega_j}{\pd q_k}\right|_x=
- \Sum_{p^{-1}(q_i)}\mathrm{Res}\left.\Omega_i\frac{\pd d\Omega_j}{\pd q_k}\right|_x =
\\
=\Sum_{p^{-1}(q_i)}\mathrm{Res}\,\left. d\Omega_i
\frac{\pd\Omega_j}{\pd q_k}\right|_x=\frac1{2\pi i}
\oint_{\d\Sigma_{\rm cut}}\left. d\Omega_i\frac{\pd\Omega_j}{\pd q_k}\right|_x - \Sum_{dx=0}\mathrm{Res}\ \left. d\Omega_i\frac{\pd\Omega_j}{\pd q_k}\right|_x =
\\
= -\Sum_{dx=0}\mathrm{Res}\,\left. d\Omega_i\frac{\pd\Omega_j}{\pd q_k}\right|_x
\ee
where we have used the fact, that singular part of $d\Omega_j$ near $q_i$ is proportional to $dx$, so the derivative $\left.\frac{\pd d\Omega_j}{\pd q_k}\right|_x$ is holomorphic, and transformed expression into the sum over all branch points $dx=0$ using the integral over the border of the cut $\Sigma$ and normalization $\oint_{A_i}d\Omega_j=0$.

Now the sum goes over the branch-points of the $x$-projection, we assume without
loss of generality, that these ramification points are simple. In the vicinity of each such point
with $(z,x)=(z^*,x^*)$ one can write (up to the terms, not giving contribution to the final formula, which is denoted by "$\approx$")
\be
x\approx x^*+a(z-z^*)^2,\,\,\,\,
z\approx z^*+\sqrt\frac{x-x^*}{a},\ \ \ \ dz \approx \frac{dx}{2\sqrt{a(x-x^*)}}
\ee
then
\be
dS\approx x^* dz,\ \ \ \ \ d\Omega_k=\left.\frac{\pd dS}{\pd q_k}\right|_z
\approx \frac{\pd x^*}{\pd q_k}dz
\\
\Omega_k\approx \frac{\pd x^*}{\pd q_k}z \approx \frac{\pd x^*}{\pd q_k}
\sqrt\frac{x-x^*}{a},
\ \ \ \
\left.\frac{\pd\Omega_j}{\pd q_k}\right|_x\approx -\frac{\frac{\pd x^*}{\pd q_j}\frac{\pd x^*}{\pd q_k}}{2\sqrt{a(x-x^*)}}
\label{der1}
\ee
and therefore
\be
\frac{\pd x^*}{\pd q_k}\approx\frac{d\Omega_k}{dz},\ \ \ \
\left.\frac{\pd \Omega_j}{\pd q_k}\right|_x\approx-
\frac{d\Omega_j}{dz}\frac{d\Omega_k}{dz}\frac{dz}{dx} =
- \frac{d\Omega_jd\Omega_k}{dx dz}
\label{der2}
\ee
so that, substituting into \rf{respre}, we finally get \rf{residue0}.
\hfill $\Box$

Note, that this formula is proven here almost in a full generality, therefore it will be
used below in all cases we need.

\subsection{AGT-correspondence and residue formulas}

The spectral curve of the large S-duality class of the quiver theories \cite{gaiotN2dual} can be written in
the form of \rf{famcur}, where the $k$-differentials $\{\phi_k\}$ are defined on the UV-curve $\Sigma_0$ -
in many cases just on Riemann sphere with marked points
 $\{z_i\} = \{0,1,\infty,q_1,q_2,q_3,\ldots\}$, where they are allowed to have some prescribed singularities. The positions of these singularities $\{q_i\}$ parameterize
the space of UV coupling-constants in the theory.

The picture is very simple in the case of $SU(2)$-quiver theory, where it has clear interpretation in
terms of two-dimensional CFT. The spectral curve equation \rf{famcur}
\be
\label{quicur}
x^2 = \phi_2(z) = \langle T(z) \rangle = \sum_{j=1}^n \left({\Delta_j\over (z-z_j)^2} + {u_j\over z-z_j}\right)
\ee
leads immediately to the residue formulas \rf{SW1} for the first derivatives of prepotentials in terms of the generating differential $dS = xdz$
\be
u_i = \res_{z=z_i} x^2 dz = \res_{z=z_i} {dSdS\over dz} =
\half\ {\rm Res}_{P_i^\pm} {dSdS\over dz} = {\d\F\over\d z_i}
\ee
Taking one more derivative (at constant $z$) one gets
\be
2x{\d x\over \d z_j} = {2\Delta_j\over (z-z_j)^3} + {u_j\over (z-z_j)^2} + {\d u_j/\d z_j\over z-z_j} +
\sum_{i\neq j}  {\d u_i/\d z_j\over z-z_i}
\ee
that is
\be
\label{2der}
{\d^2\F\over\d z_j} = 2\res_{z=z_j} x{\d x\over \d z_j}dz = 2\res_{z=z_j} {dSd\Omega_j\over dz}
\\
{\d^2\F\over\d z_i\d z_j} = 2\res_{z=z_i} x{\d x\over \d z_j}dz = 2\res_{z=z_i} {dSd\Omega_j\over dz},\ \ \ \ i\neq j
\ee
where
\be
d\Omega_j = {\d x\over \d z_j}dz = \nabla^z_{\d\over\d z_i} xdz = \nabla^z_{\d\over\d z_i} dS
\ee
and the derivatives
\be
{\d  u_k\over\d z_i} = \left.{\d u_k\over\d z_i}\right|_{\bf a}
\ee
are taken at constant ${\bf a}$-periods ${\bf a} = {1\over 2\pi i}\oint_{\bf A} xdz$ of the generating differential or some fixed choice of the cycles $\{ A_j\}$ on the cover.

Notice, that these formulas are not all independent due to constraints, coming from the regularity condition at $z=\infty$
\be
\label{3constr}
\sum_{j=1}^n u_j=0,\ \ \ \ \sum_{j=1}^n (z_ju_j+\Delta_j)
= 0,\ \ \ \ \
\sum_{j=1}^n (z_j^2u_j+2z_j\Delta_j)
= 0
\ee
Consider now the reparameterization in the space of bare coupling induced by
$z\rightarrow \omega(z)$, which
can be conveniently encoded by $d\omega = dz/f(z)$.
Then, the first derivatives of deformed prepotential
$\F_f$ are given by
\be
f(z_i)\frac{\d\mathcal{F}_f}{\d z_i}= \res_{z=z_i}\frac{dSdS}{dz/f(z)} = \res_{z=z_i}f(z)x^2dz
\ee
Calculating the residue in the r.h.s. using \rf{quicur} one finds, that
$$
f(z_i)\frac{\pd\mathcal{F}_f}{\pd z_i}=f(z_i)u_i+f'(z_i)\Delta_i
$$
and this corresponds to the transformation
\be
\mathcal{F}_f=\mathcal{ F}+\Sum_{i=1}^n\Delta_i\log f(z_i)
\ee
which does not change the derivatives of prepotential over the period $\mathbf{a}$-variables.

Notice that the residue formula is also true for an arbitrary $d\omega={dz\over f(z)}$:
\be
\frac{\d^3\mathcal{F}_f}{\d \omega(z_i)\d \omega(z_j)\d \omega(z_k)}=\Sum_{d(xf(z))=0}\frac{d\Omega^f_id\Omega^f_jd\Omega^f_k}{d(xf(z))d\omega}
\ee
where $d\Omega^f_i=\left.f(z_i)\frac{\d dS}{\d z_i}\right|_z$. It is clear, since the proof of sect.~\ref{ss:res} can be rewritten
literally for the differential $dz/f(z)$ and the function $xf(z)$.

Similarly one can consider the change of couplings, corresponding
$$
(z_1,\ldots,z_n)\rightarrow (q_1,\ldots,q_{n-3},1,0,\infty)
$$
In particular, for $f(z) = {(z-z_n)(z-z_{n-1})\over z_n-z_{n-1}}$ with some fixed $(z_n,z_{n-1},z_{n-2})$ one gets
\be
\label{qdegen}
z_j{\d\F_f\over\d z_j} = \res_{z_j}{dS\over d\omega}dS   = \res_{z_j}\ {x^2 dz\over d\omega/dz} =
\\
= {2z_j-z_n-z_{n-1}\over z_n-z_{n-1}}\Delta_j+{(z_j-z_n)(z_j-z_{n-1})\over z_n-z_{n-1}}u_j
\ \ \ \ j=1,\ldots,n-3
\ee
where $d\omega={dz\over f(z)} = {dz\over z-z_n} - {dz\over z-z_{n-1}}$. In this way one can
easily reproduce all original formulas from \cite{AMtau}.

\setcounter{equation}0
\section{Weak-coupling expansions of the prepotentials
\label{ss:wcexp}}

In this section we propose the techniques of the weak-coupling analysis of the quiver gauge theories, based on applications of the residue formulas. For the perturbative prepotentials - instead of computation of
the period matrices of degenerate curves - one can just compute the residues of certain one-forms, which can be projected from the SW curve $\Sigma$ to the UV Gaiotto curve $\Sigma_0$. This procedure has
been applied to the
computations of the dependence of perturbative prepotentials over vacuum condensates in \cite{MaMi}, and we extend it here to the tau-functions of quiver gauge theories as functions of the bare couplings.

The dependence of perturbative prepotentials over the UV couplings is rather simple and can be
directly compared with the one-loop calculations in corresponding supersymmetric
quantum field theories. However, the application of
our methods can be immediately extended to compute the whole weak-coupling expansion of a
prepotential~\footnote{The neighborhoods of degenerate curves in the moduli space are usually called as the weak-coupling regions, this is true indeed in many cases - the corresponding theory has a Lagrangian description, and the whole weak-coupling expansion can be recovered from the instanton calculus. }.
One can apply for these purposes the differential equations, obtained from the residue formulas
\rf{residue0} for the third $\mathbf{q}$-derivatives, and expressing them in terms of the first
\rf{SW1} and second
\rf{2ndq} derivatives of the same prepotentials. Equivalently, one can compute the power corrections in
bare couplings to the perturbative prepotentials by study of the $\mathbf{q}$-expansions of
the period integrals, which define the integration constants for these equations. These
power corrections exactly correspond to the instanton expansions of the quiver gauge theories,
but also go beyond this case, when the latter cannot be defined \cite{nepes}. As an example,
we compute the expansion for the case of sicilian quiver with three $SU(2)$ groups, which
will be also used later for discussion of the constrained or Zamolodchikov's case.

\subsection{Methods for the weak-coupling expansion}

An effective solution of the equations (\ref{SW0}) is generally not so easy due to the complicated geometry of the spectral curve (see e.g. \cite{DHoPho}). Fortunately, in the vicinity of particular points in the moduli space, where spectral curve degenerates, one can find the series expansions of the prepotential. Here we describe two different but closely related methods of such calculation.

\underline{Method I:}
\begin{itemize}
\item Parameterize a spectral curve through $\frac{\pd\mathcal{F}}{\pd q_i}$ using (\ref{SW1});
\item Substitute an expansion $\mathcal{F}=\mathbf{A}\log \mathbf{q}+\Sum_{\mathbf{k}>0} c_\mathbf{k}\mathbf{q}^\mathbf{k}$ into the first half of equations in (\ref{SW0}) and solve obtained equations iteratively;
\item Recover the $\mathbf{q}$-independent part of the prepotential using (\ref{residue}) for the $\mathbf{a}$-derivatives.
\end{itemize}

\underline{Method II:}

\begin{itemize}
\item Derive the non-linear differential equation for the prepotential as the function of $\mathbf{q}$ using (\ref{residue}), it expresses the third derivatives through the first (coefficients of
    the equation for the curve) and second (coefficients in the expressions for the differentials
    $d\Omega$);
\item Solve the first equation of (\ref{SW0}) in the degenerate limit, and recover the term $\mathbf{A}\log\mathbf{q}$;
\item Substitute an expansion $\mathcal{F}=\mathbf{A}\log \mathbf{q}+\Sum_{\mathbf{k}>0} c_\mathbf{k}\mathbf{q}^\mathbf{k}$ into obtained differential equation and solve it iteratively;
\item Recover again the $\mathbf{q}$-independent part of the prepotential using (\ref{residue}) for the $\mathbf{a}$-derivatives.
\end{itemize}

Strictly speaking, the first method is just a modification of the second, since it uses directly
the period integrals, which play the role of the integrals of motion for the differential equations we use in the second method of computation. In what follows we use both of them, dependently on particular example, to save the efforts.

\subsection{Warm-up examples}

\paragraph{Original SW theory}


Let us start with the integrable system from the first work of Seiberg and Witten \cite{SW}
\be
y^2 = (x^2-\Lambda^4)(x-u),\ \ \ \ dS=\sqrt\frac{2(u-x)}{x^2-\Lambda^4}dx
\ee
Introduce $\lambda_1$ and $\lambda_2$ such, that $\Lambda^2=\lambda_1-\lambda_2$ and
\be
dS=\sqrt\frac{2\left(\frac{\lambda_1+\lambda_2}2+u-x\right)}{(x-\lambda_1)(x-\lambda_2)}dx
\ee
(one can always put $\lambda_c=\lambda_1+\lambda_2=0$ at the end), then in follows from \rf{SW1}, that
\be
\frac{\pd\mathcal{F}}{\pd\lambda_1}={\rm Res}_{\lambda_1}\frac{2(\frac{\lambda_1+\lambda_2}2+u-x)}
{(x-\lambda_1)(x-\lambda_2)}=
\frac{\lambda_2-\lambda_1+2u}{\lambda_1-\lambda_2}
\\
\frac{\pd\mathcal{F}}{\pd\lambda_2}={\rm Res}_{\lambda_2}\frac{2(\frac{\lambda_1+\lambda_2}2+u-x)}
{(x-\lambda_1)(x-\lambda_2)}=\frac{\lambda_1-\lambda_2+2u}{\lambda_2-\lambda_1}
\ee
Since $\frac\pd{\pd \lambda_1}=\frac\pd{\pd\Lambda^2}+\frac\pd{\pd\lambda_c}$,
$\frac\pd{\pd \lambda_2}=-\frac\pd{\pd\Lambda^2}+\frac\pd{\pd\lambda_c}$, one gets
therefore $\frac{\pd\mathcal{F}}{\pd\lambda_c}=1$ and
\be
\Lambda\frac{\pd\mathcal{F}}{\pd\Lambda}=4u
\ee
Substituting
\be
d\Omega=\frac{\pd dS}{\pd\Lambda^2}=\left(\frac{\Lambda^2}{x^2-\Lambda^4}+\frac1{4\Lambda}\frac{\pd u}{\pd\Lambda}\frac1{u-x}\right)ydx
\ee
into the residue formula \rf{residue} we get
\be
\frac{\pd^3\mathcal{F}}{(\pd\Lambda^2)^3}=-2\Sum_{x\in\{\pm\Lambda^2,u\}}\mathrm{Res}\frac{(d\Omega)^3}{dx dy}=\Sum_{x\in\{\pm\Lambda^2,u\}}
\mathrm{Res}\left(\frac{2\Lambda^2}{x^2-\Lambda^4}+\frac1{2\Lambda}\frac{\pd u}{\pd\Lambda}\frac1{u-x}\right)^3\frac{dx}{(2/y^2)'}
\ee
and computing the residue we get an equation for the prepotential:
\be
\boxed{4\Lambda^2\left(\left(\frac{\pd\mathcal{F}}{\pd\Lambda}\right)^2-16\Lambda^2\right)\frac{\pd^3\mathcal{F}}{\pd\Lambda^3}+
\left(\Lambda\frac{\pd^2\mathcal{F}}{
\pd\Lambda^2}-\frac{\pd\mathcal{F}}{\pd\Lambda}\right)^3=0}
\ee
It is certainly well-known (see, e.g. \cite{Matone}) and even equivalent in this case to the hypergeometric differential equation for the inverse function. The only important for us thing is that it comes also from
the residue formula \rf{residue} and allows to determine immediately the weak-coupling expansion of the prepotential, substituting an ansatz
$\mathcal{F}=2 a^2\log\Lambda+\Sum_{k=1}^\infty c_k
\Lambda^{4k} a^{2-4k}$ and solving the algebraic equations for the coefficients $c_k$ with the result
\be
\mathcal{F}=-2a^2\log a+2a^2\log\Lambda+\frac{\Lambda^4}{4a^2}+\frac{5\Lambda^8}{128a^6}+\frac{3\Lambda^{12}}{128 a^{10}}+\ldots
\ee
Notice only, that everywhere in this example we used the original normalization of \cite{SW} for the
period $a$, which corresponds to the mass of $W$-boson.

\paragraph{Conformal $SU(2)$ theory}

Now, again for the illustration purposes, consider the $SU(2)$ theory with four massless flavors, corresponding to sphere with $n=4$ marked points. Equation \rf{quicur} acquires the form
\be
\label{confqcd}
x^2=\frac{u}{z(z-1)(z-q)}=\frac{q(q-1)\frac{\d\mathcal{F}}{\d q}}{z(z-1)(z-q)}
\ee
since
\be
\frac{\d\mathcal{F}}{\d q}=\res_{z=q} x^2dz = \frac{u}{q(q-1)}
\ee
The residue formula
\be
\left(\frac{\d\mathcal{F}}{\d q}\right)^2
\frac{\d^3\mathcal{F}}{\d q^3}=\frac12{\rm Res}_q\left(\left(\frac1{z-q}+\frac1q+\frac1{q-1}\right)\frac{\d\mathcal{F}}{\d q}+
\frac{\d^2\mathcal{F}}{\d q^2}\right)^3\frac{q(q-1) (dz)^2}{d\left(z(z-1)(z-q)\right)}
\ee
gives rise to the differential equation
\be
\frac{\d\mathcal{F}}{\d q}\frac{\d^3\mathcal{F}}{\d q^3}=\frac32\left(\frac{\d^2\mathcal{F}}{\d q^2}\right)^2+
\frac{1-q+q^2}{2q^2(q-1)^2}\left(\frac{\d\mathcal{F}}{\d q}\right)^2
\ee
which can be rewritten in the form
\be
\label{schw}
\{\mathcal{F},q\}=\frac{1-q+q^2}{2q^2(q-1)^2}
\ee
where $\{,\}$ stays for the Schwarzian derivative. The general solution is
\be
\mathcal{F}=\frac{A K(1-q)+B K(q)}{C K(1-q)+D K(q)},\ \ \ \ \left(\begin{array}{cc}
                                                                    A & B \\
                                                                    C & D
                                                                  \end{array}\right)\in PSL_2(\mathbb{C})
\ee
which comes from the fact, that $K(q) = \Int_0^q\frac{dz}{\sqrt{z(z-1)(z-q)}}$ and $K(1-q) =\Int_q^1\frac{dz}{\sqrt{-z(z-1)(z-q)}}$
form the basis of solutions to $f''(q)+T(q)f(q)=0$, with $T(q) = \frac{1-q+q^2}{2q^2(q-1)^2}$.

To fix the physical solution we should impose, that $\F =a^2 \log q + \ldots$, which
gives
\be
\label{prepn4}
\F = i\pi a^2 \tau( q) = -\pi a^2\frac{K(1-q)}{K(q)}=
\\
= a^2\left(\log q-\log 16 +\frac q2+\frac{13 q^2}{64}+\frac{23q^3}{192}+
\frac{2701 q^4}{32768}+\ldots\right)
\ee
This is just one more way to get the non-perturbative renormalization of coupling in the conformal theory with the vanishing beta-function (cf. with \cite{ZAGTKl,ZAGT,Pogh,Egumar}).

\subsection{Quiver gauge theory and S-duality class}

\paragraph{$SU(2)\times SU(2)$ linear quiver
\label{ss:quiver}}


Let us turn to the quiver gauge theories and consider, first, the $SU(2)\times SU(2)$ gauge quiver with four massless fundamentals and one bi-fundamental multiplet. We consider it in the limit $\epsilon\to 0$
after reparameterization $q_1=\epsilon Q_1$, $q_2=1-\epsilon Q_2$, here the parameter $\epsilon$ plays the role of degree-counting
variable, so we will put $\epsilon=1$ in the final answer. The spectral curve equation \rf{quicur} now reads
\be
x^2=\frac{q_1(q_1-1)\frac{\pd\mathcal{F}}{\pd q_1}}{z(z-1)(z-q_1)}+\frac{q_2(q_2-1)\frac{\pd\mathcal{F}}{\pd q_2}}{z(z-1)(z-q_2)} =
\\
= \frac{(1-z)F_1(1-\epsilon Q_1)+zF_2(1-\epsilon Q_2)+\epsilon^2Q_1Q_2(F_1+F_2)
}{z(\epsilon Q_1-z)(z-1)(1-\epsilon Q_2-z)}
\label{LQcurve}
\ee
where we parameterized the curve by
$F_1=q_1\frac{\pd\mathcal{F}}{\pd q_1}=\frac{\pd\mathcal{F}}{\pd\log Q_1}$, and $F_2=(q_2-1)\frac{\pd\mathcal{F}}{\pd q_2}=\frac{\pd\mathcal{F}}{\pd\log Q_2}$.

Now we can compute the periods $a_i=\frac1{2\pi i}\Oint_{A_i} xdz$, $i=1,2$, expanding
these integrals into the series. Namely,
\be
\label{a1exp}
a_1=\frac{\sqrt{F_1}}\pi\Int_0^{\epsilon Q_1}\sqrt\frac{(1-z)(1-\epsilon Q_1)+z\frac{F_2}{F_1}(1-\epsilon Q_2)
+\epsilon^2Q_1Q_2(1+\frac{F_2}{F_1})}{(1-z)(1-\epsilon Q_2-z)}\frac{dz}{\sqrt{z(\epsilon Q_1-z)}} =
\\
= \frac{\sqrt{F_1}}\pi\Int_0^{\epsilon Q_1}\left(1+\sum_{k=1}^\infty f_{1,k}
 z^k\right)\frac{dz}{\sqrt{z(\epsilon Q_1-z)}} =\\=
\sqrt{F_1}\left(1+\Sum_{k=1}^\infty \epsilon^k Q_1^k\frac{(2n-1)!!}{(2n)!!}f_{1,k}(\epsilon Q_1,\epsilon Q_2,F_2/F_1)\right)
\ee
where the integrals were computed using
\be
\label{intcalc}
\frac1\pi\Int_0^{\epsilon Q_1}\frac{z^n dz}{\sqrt{z(\epsilon Q_1-z)}}=\epsilon^n Q_1^n\frac{\Gamma(n+\frac12)}{\sqrt\pi\Gamma(n+1)}=
\epsilon^n Q_1^n\frac{(2n-1)!!}{(2n)!!}
\ee
The same should be done with the $A_2$-period
\be
\label{a2exp}
a_2=\frac{\sqrt{F_2}}{\pi}\Int_{1-\epsilon Q_2}^1\left(1+\Sum_{k=1}^\infty f_{2,k}(\epsilon Q_1,\epsilon Q_2,F_1/F_2)(z-1)^k\right)\frac{dz}{\sqrt{(1-z)(z-1+\epsilon Q_2)}}=
\\
=\sqrt{F_2}\left(1+\Sum_{k=1}^\infty (-1)^k\epsilon^k Q_2^k\frac{(2n-1)!!}{(2n)!!}f_{2,k}(\epsilon Q_1,\epsilon Q_2,F_1/F_2)\right)
\ee
Explicitly for the expansions \rf{a1exp} and \rf{a2exp}, one gets
\be
a_1=\sqrt{F_1}-\frac{Q_1(F_1+F_2)}{4\sqrt{F_1}}\epsilon-\frac{Q_1^2(7F_1^2+14F_1F_2+3F_2^2)}{64 F_1^{3/2}}\epsilon^2-
\\
-\frac{Q_1^2(17F_1^3Q_1+51F_1^2F_2Q_1+23F_1F_2^2Q_1+5F_2^3Q_1+16F_1^2F_2Q_2)}{256F_1^{5/2}}\epsilon^3+\ldots
\\
a_2=\sqrt{F_2}-\frac{Q_2(F_1+F_2)}{4\sqrt{F_2}}\epsilon-\frac{Q_2^2(7F_2^2+14F_2F_1+3F_1^2)}{64 F_2^{3/2}}\epsilon^2-
\\
-\frac{Q_2^2(17F_2^3Q_2+51F_2^2F_1Q_2+23F_2F_1^2Q_2+5F_1^3Q_2+16F_2^2F_1Q_1)}{256F_2^{5/2}}\epsilon^3+\ldots
\ee
Substituting here $F_1=a_1^2+\Sum_{k=1}^\infty F_{1,k}\epsilon^k$, $F_2=a_2^2+\Sum_{k=1}^\infty F_{2,k}\epsilon^k$
and inverting these equations, one can check, in particular, that $Q_1\frac{\pd F_2}{\pd Q_1}=Q_2\frac{\pd F_1}{\pd Q_2}$, and get the final expression for the expansion of the prepotential. It reads (after the substitution $\epsilon=1$)
\be
\label{2su2}
\mathcal{F}(\mathbf{a},\mathbf{q})= \F_{\rm pert}(\mathbf{a}) +
a_1^2 \log Q_1+a_2^2 \log Q_2+\frac{a_1^2+a_2^2}2(Q_1+Q_2)+
\\
+\frac{13 a_1^4+18 a_1^2a_2^2+a_2^4}{64 a_1^2}Q_1^2+
\frac{a_1^2+a_2^2}4 Q_1Q_2+\frac{13 a_2^4+18 a_1^2a_2^2+a_1^4}{64 a_2^2}Q_2^2+\\+
\frac{23a_1^4+38a_1^2a_2^2+3a_2^4}{192a_1^2}Q_1^3+\frac{13 a_1^4+18 a_1^2a_2^2+a_2^4}{64 a_1^2}Q_1^2Q_2+
\\
+\frac{13 a_2^4+18 a_1^2a_2^2+a_1^4}{64 a_2^2}Q_1Q_2^2+
\frac{23a_2^4+38a_1^2a_2^2+3a_1^4}{192a_2^2}Q_2^3+\\+
\frac{2701a_1^8+5028a_1^6a_2^2+470a_1^4a_2^4-12a_1^2a_2^6+5a_2^8}{32768a_1^6}Q_1^4+\frac{23a_1^4+38a_1^2a_2^2+3a_2^4}{128a_1^2}Q_1^3Q_2+
\\
+\frac{17a_1^6+343a_1^4a_2^2+343a_1^2a_2^4+17a_2^6}{1024a_1^2a_2^2}Q_1^2Q_2^2+\ldots
\ee
Note that the coefficients $\frac12, \frac{13}{64}, \frac{23}{192}, \frac{2701}{32768}$ in \rf{2su2} are the coefficients of expansion
$-\pi\frac{K(1-q)}{K(q)}$, or of the prepotential \rf{prepn4} for a single $SU(2)$ group.

To fix the perturbative part one can apply the residue formula \rf{residue} for the $\mathbf{a}$-variables
\be
\frac{\pd^3\mathcal{F}_{\rm pert}}{\pd a_1^3}=-2\Sum_{dz=0}\mathrm{res\,}\left(\frac{\pd\log x}{\pd a_1}\right)^3x^2\frac{dz}{(\log x)'}=
\\
= -2\Sum_{dz=0}\mathrm{res\,}\left(\frac12\frac{\pd\alpha}{\pd a_i}-\frac12\frac{\pd v}{\pd a_1}\frac1{z-v}\right)^3\frac{
2\alpha(z-v)dz}{\prod(z-z_i)(\frac1{z-v}-\sum\frac1{z-z_i})}=\\
=-2\mathrm{res}_{z=v}\frac{dz}{z-v}\left(\frac12(z-v)\frac{\pd\alpha}{\pd a_i}-\frac12\frac{\pd v}{\pd a_1}\right)^3\frac{2\alpha}{
\prod(z-z_i)(1-\sum\frac{z-v}{z-z_i})}=
\\
= \frac12\left(\frac{\pd v}{\pd a_1}\right)^3\frac{\alpha}{\prod(v-z_i)}
\ee
on the degenerated curve (\ref{LQcurve})
\be
x^2=\frac{(1-z)a_1^2+za_2^2}{z(z-\epsilon Q_1)(z-1)(z-1+\epsilon Q_2)}=\frac{\alpha(z-v)}{\prod(z-z_i)}
\ee
where we have substituted $\epsilon\to 0$ in the numerator, i.e. $F_1=a_1^2$, $F_2=a_2^2$. In the limit $\epsilon\to 0$: $v=\frac{a_1^2}{a_1^2-a_2^2}$, $\alpha=a_2^2-a_1^2$, i.e.
\be
\frac{\pd v}{\pd a_1}=-\frac{2a_1a_2^2}{(a_1^2-a_2^2)^2},\,\,\,\,\,\prod(v-z_i)=\frac{a_1^4a_2^4}{(a_1^2-a_2^2)^4}
\ee
and we obtain
\be
\frac{\pd^3\mathcal{F}_{\rm pert}}{\pd a_1^3} =
\frac2{a_1-a_2}+\frac2{a_1+a_2}-\frac4{a_1}
\ee
which gives for the $\mathbf{q}$-independent part
\be
\mathcal{F}_{\rm pert}=(a_1-a_2)^2\log (a_1-a_2)+(a_1+a_2)^2\log (a_1+a_2)-2a_1^2\log a_1-2a_2^2\log a_2
\ee

\paragraph{$SU(2)\times SU(2)\times SU(2)$ sicilian quiver}

For the $SU(2)\times SU(2)\times SU(2)$ theory the curve \rf{quicur} is parameterized as
\be
\label{siquicu}
x^2=\frac{Q_1(\epsilon Q_1-1)\frac{\d\mathcal{F}}{\d Q_1}}{z(z-\epsilon Q_1)(z-1)}+
\frac{(1-\epsilon Q_2)Q_2\frac{\d\mathcal{F}}{\d Q_2}}
{z(z-1+\epsilon Q_2)(z-1)}+\frac{(1-\frac1{\epsilon Q_3})Q_3\frac{\d\mathcal{F}}{\d Q_3}}{z(z-1)(z-\frac1{\epsilon Q_3})}
\ee
We have chosen parametrization $q_1=\epsilon Q_1$, $q_2=1-\epsilon Q_2$, $q_3={1\over\epsilon Q_3}$ in the space of UV couplings to make it convenient for the
computations in the weak-coupling phase for sicilian quiver with massless fundamental and
(half-) tri-fundamental multiplets. This parametrization is adjusted to particular degeneration of the UV curve in such a way, that one has a single central component (corresponding to the trifundamental) connected to three another components (each corresponding to a pair of fundamentals)~\footnote{
Notice, that it is essentially different from parameterization $q_1=\epsilon^3 Q_1Q_2Q_3$, $q_2=\epsilon^2Q_2Q_3$, $q_3=\epsilon Q_3$, convenient for the computations for
the linear quiver. It is easy to verify, e.g. that for these two choices of parameterization
the $\epsilon\to 0$ limit is consistent with fixing homology on the curve, corresponding to particular perturbative
phase of gauge theory, which in its turn is manifested by singularities in the period matrices
and expansion of the perturbative prepotential.}.

Denote again $F_i=Q_i\frac{\pd\mathcal{F}}{\pd Q_i}$, $i=1,2,3$, and solve equations for the $A$-periods of the type \rf{a1exp}, \rf{a2exp}. Now we need to compute one more integral
$\Int_{\frac1{\epsilon Q_3}}^\infty\frac{z^{-k-1}dz}{\sqrt{\epsilon Q_3 z-1}}$, corresponding to the third
$A$-period, which is calculated again, using formula similar to \rf{intcalc}. Just the same procedure as
in the case of two gauge groups leads now in a straightforward way to the answer
\be
\label{trifundexp}
\mathcal{F}=a_1^2\log Q_1+a_2^2\log Q_2+a_3^2\log Q_3-2a_1^2\log a_1-2a_2^2\log a_2-2a_3^2\log a_3+
\\
+\frac12\Sum_{\epsilon,\epsilon'=\pm}
 (a_1+\epsilon a_2+\epsilon' a_3)^2\log(a_1+\epsilon a_2+\epsilon' a_3)+\\+
\frac{a_1^2+a_2^2-a_3^2}2Q_1+\frac{a_1^2+a_2^2-a_3^2}2Q_2+\frac{-a_1^2+a_2^2+a_3^2}2Q_3+
\\
 +\frac{a_1^2+a_2^2-a_3^2}4Q_1Q_2+\frac{a_1^2-a_2^2-a_3^2}4Q_2Q_3+\frac{a_1^2-a_2^2+a_3^2}4Q_1Q_3+
 \\+\frac{13 a_1^4+18a_1^2a_2^2-14a_1^2a_3^2+a_2^4-2a_2^2a_3^2+a_3^4}{64a_1^2}Q_1^2+\\+
 \frac{a_1^4+13a_2^4-14a_2^2a_3^2+a_3^4+18a_1^2a_2^2-2a_1^2a_3^2}{64a_2^2}Q_2^2+
 \\
 +\frac{a_1^4+a_2^4+18a_2^2a_3^2+13a_3^4-2a_1^2a_2^2-14a_1^2a_3^2}{64a_3^2}Q_3^2+...
 \ee
Let us stress here, that the result in this case, where the standard methods \cite{Nek} of the instanton calculus
are not applicable directly \cite{nepes}, is obtained from the study of gauge-theory tau-functions exactly in the same way as for the theories, where the weak-coupling expansion is saturated by the instanton configurations. This allows us to hope for a direct application of our methods for the S-duality class of generic
$SU(N)$ quiver gauge theories, which can shed light to the physical properties of supersymmetric gauge
theories, which do not even have a Lagrangian formulation.

\subsection{Mass-deformed theory and quasiclassical conformal block
\label{ss:masses}}


For the $n=4$ massless $SU(2)$ theory the prepotential is given by quadratic expression \rf{prepn4}.
Consider now its simplest deformation, when two flavors receives an opposite masses, e.g.
$\Delta_0=m^2$, $\Delta_1=\Delta_q=\Delta_\infty=0$, with the curve \rf{quicur} for this case
\be
x^2=\frac{zq(q-1)\mathcal{F}'-(z-q)m^2}{z^2(z-1)(z-q)}
\ee
where $\mathcal{F}' = \frac{\pd\mathcal{F}}{\pd q}$. The residue formula \rf{residue0} gives now the differential equation
\be
\label{difeq1m}
\mathcal{F}'''+
\\
+ \frac{
m^4\mathcal{F}'[3q(2-3q)\mathcal{F}''+2(1-3q)\mathcal{F}']-[3q^4(q-1)^2\mathcal{F}'^2\mathcal{F}''^2
+q^2(q^2-q+1)\mathcal{F}'^4]}{2 q^2(q-1)
\mathcal{F}'[q^2(q-1)\mathcal{F}'^2+m^2q(q-2)\mathcal{F}'-m^4]}-\\-
\frac{m^2[(q-1)^2q^2\mathcal{F}''^3+6q(q-1)^2\mathcal{F}'
\mathcal{F}''^2+3(q^2+q-1)\mathcal{F}''\mathcal{F}'^2+(3+2q)\mathcal{F}'^3]}{2(q-1)
\mathcal{F}'[q^2(q-1)\mathcal{F}'^2+m^2q(q-2)\mathcal{F}'-m^4]}=0
\ee
which can be solved, using the anzatz $\mathcal{F}=(a^2-m^2)\log q+\Sum_{k=1}^\infty c_kq^k$, giving
rise to expansion
\be
\label{1mass}
\mathcal{F}=\mathcal{F}_{\rm pert}(a;m)+(a^2-m^2)\log q+\frac{a^2-m^2}2q+
\\
+\frac{13a^4-14a^2m^2+m^4}{64a^2}q^2+
\frac{23a^4-26a^2m^2+3m^4}{192a^2}q^3+
\\
+\frac{2701a^8-3164a^6m^2+470a^4m^4-12a^2m^6+5m^8}{32768a^6}q^4 + \ldots
\ee
Notice, that this prepotential is directly related to the corresponding expression \rf{2su2}
in the massless $SU(2)\times SU(2)$ theory
(we compare $\mathcal{F}(a,m,q)$ for a single massive flavor theory with  the massless prepotential $\mathcal{F}(a_1,a_2,q_1,1-Q_2)$ for two gauge groups \rf{2su2} in the limit $Q_2=0$, $a_2=m$).
Their difference
\be
\left[\left.\mathcal{F}(a,\tilde{a},q,1-Q)\right|_{\tilde{a}=m}-m^2\log Q\right]_{Q=0}-
\mathcal{F}(a,m,q)=-m^2\log (1-q)
\ee
is just a $U(1)$-factor, commonly arising in the context of the
AGT correspondence \cite{AGT}. The q-independent term in \rf{1mass}
\be
\label{pertm0}
\mathcal{F}_{\rm pert}(a;m)=(a-m)^2\log (a-m)+(a+m)^2\log (a+m)-2a^2\log a
\ee
is restored in standard way from residue formula on degenerate curve, and it vanishes in the limit $m\to 0$.

Now let us add more massive deformations for a single $SU(2)$ and consider generic four-point function
\be
\label{eQ4}
x^2=\frac{(q-1)q\mathcal{F}'}{z(z-q)(z-1)}+\frac{\Delta_0}{z^2}+\frac{\Delta_1}{(z-1)^2}+
\frac{\Delta_q}{(z-q)^2}-
\frac{\Delta_0+\Delta_1+\Delta_q-\Delta_\infty}{z(z-1)}=
\\
=\frac{Q_4(z)}{z^2(z-1)^2(z-q)^2} = \phi_2(z)
\ee
where four $\Delta=\Delta(\mathbf{m})$ are quadratic functions 
of the fundamental masses only.
Denote $q\mathcal{F}'|_{q=0}=A$ and look first for the solution in the weak-coupling region
$q\rightarrow 0$. One has
\be
Q_4^{(0)}(z) = \left.Q_4(z)\right|_{q=0}=z^2\left(\Delta_\infty z^2+(-A-\Delta_0+\Delta_1-\Delta_q-\Delta_\infty)z+(A+\Delta_0+\Delta_q)\right)
\\
\frac{\pd x}{\pd a}= \frac1{2\sqrt{Q_4(z)}}\frac{\pd Q(z)}{\pd a}\frac{1}{z(z-1)(z-q)}\
\stackreb{q\to 0}{\approx}\ -\frac{\frac{\pd A}{\pd a}}{2\sqrt{Q_4^{(0)}(z)}}
\ee
and from the normalization of the holomorphic differential
\be
1\approx-\frac1{4\pi i}\frac{\pd A}{\pd a}\Oint_A\frac{dz}{\sqrt{Q_4^{(0)}(z)}}
\approx-\frac12\frac{\pd A}{\pd a}\frac1{\sqrt{A+\Delta_0+\Delta_q}}
\ee
which gives $A=a^2-\Delta_0-\Delta_q$, i.e. the leading exponent for $q\to 0$ expansion of the four-point conformal block on sphere.

The differential equation is obtained similarly to \rf{difeq1m}, though it requires for generic
massive deformation some additional efforts - to sum in the residue formula
\be
\mathcal{F}''' = -\res_{z=q,Q_4(z)=0} {(\phi_2')^3dz\over 2\phi_2\frac{d\phi_2}{dz}} =
-\res_{z=q,Q_4(z)=0}\,\frac{T(z)dz}{Q_4(z)S(z)}
\ee
over the unknown roots of the polynomial $Q_4(z)$ in the equation \rf{eQ4} in addition to the
fourth-order pole at $S(z)\ \stackreb{z\to q}{\sim}\ (z-q)^4$.
Calculating the sum over the zeroes of a polynomial $Q_k(z)=\prod_{i=1}^k(z-\lambda_i)$
\be
\label{sumresdiscr}
\Sum_{i=1}^k\frac{T(\lambda_i)}{S(\lambda_i)Q'(\lambda_i)}=
\frac{\Sum_{i=1}^k T(\lambda_i)\prod_{j\neq i}(Q'(\lambda_j)S(\lambda_j))}{\prod_{i=1}^k S(\lambda_i)Q'(\lambda_i)}=
\frac{\Sum_{i=1}^k T(\lambda_i)\prod_{j\neq i}(Q'(\lambda_j)S(\lambda_j))}{\mathrm{R}\{S,Q\}\mathrm{D}\{Q\}}
\ee
where $\mathrm{R}\{S,Q\}$ is the resultant and $\mathrm{D}\{Q\}$ stays for the discriminant,
one gets some rational symmetric function of the roots of $Q_4(z)$.

Once the differential equation was derived, we substitute the perturbative expansion\par\noindent $\mathcal{F}=A\log q+\Sum_{i=1}^\infty c_iq^i$
and obtain  an answer for the prepotential (here the result for $\Delta_0=\Delta_q=0$ and arbitrary $\Delta_1=m_1^2$ and $\Delta_\infty=m_\infty^2$ is presented~\footnote{
From the physical point of view, as in \rf{pertm0}, the perturbative part is a result of partial cancelation
\be
\mathcal{F}_{\rm pert} = \half\Sum_{\epsilon,\epsilon'}\left[(a+\epsilon m_0+\epsilon' m_q)^2\log(a+\epsilon m_0+\epsilon' m_q)+(a+\epsilon m_1+\epsilon' m_\infty)^2\log(a+\epsilon m_1+\epsilon' m_\infty)\right]-
\\
- 4a^2\log a
\ee
between the contribution of massless fundamental and vector multiplets at $m_0=m_q=0$.
}):
\be
\mathcal{F}=a^2\log q+\half\Sum_{\epsilon,\epsilon'}(a+\epsilon m_1+\epsilon' m_\infty)^2\log(a+\epsilon m_1+\epsilon' m_\infty)-2a^2\log a+
\\+\frac{a^2+m_1^2-m_\infty^2}2q+\frac{13a^4+18a^2m_1^2-14a^2m_\infty^2+m_1^4+m_\infty^4-
2m_1^2m_\infty^2}{64a^2}q^2+\\+
\frac{207a^6+a^4(334m_1^2-226m_\infty^2)+a^2(43m_1^4-54m_1^2m_\infty^2+11m_\infty^4)-
8(m_1^2-m_\infty^2)^3}{1728a^4}q^3+
\\
+ O(q^4)
\label{block1}
\ee
It is instructive to compare this result with the expansion for the quasiclassical conformal block from \cite{NekZam}, depending on intermediate dimension $\Delta=a^2$ in addition to the external dimensions. For two non-vanishing external dimensions, as in \rf{block1}, the formula for quasiclassical conformal block gives
\be
f(q)=a^2\log q+\frac{a^2+m_1^2-m_\infty^2}2q+
\left(\frac{a^2+m_1^2-m_\infty^2}4+ \right.
\\
\left.+ \frac{a^4+2a^2(m_1^2+m_\infty^2)-
3(m_1^2-m_\infty^2)^2}{64(a^2+\frac34)}-\frac{a^4-(m_1^2-m_\infty^2)^2}{16a^2}\right)q^2+\ldots
\label{block2}\ee
and the single mass case is easily reproduced by $m_\infty\mapsto m\neq 0$, $m_1\mapsto 0$. It is easy to see,
that expressions (\ref{block1}) and (\ref{block2}) literally coincide in the SW limit for conformal blocks, when all dimensions $\Delta\rightarrow\infty$, including intermediate, simultaneously. Then almost all terms remain intact except for $\frac{3}{4\Delta}\rightarrow 0$, and the correction in denominator from the inverse Shapovalov form disappear.

In convenient parametrization for two-dimensional conformal theory $\epsilon_1=
bg$, $\epsilon_2=\frac gb$, the central charge is $c=1+6\frac{(\epsilon_1+\epsilon_2)^2}{\epsilon_1\epsilon_2}=1+6(b+\frac1b)^2$, and for conformal dimensions one can write $\Delta(\alpha)=
\frac{(\epsilon_1+\epsilon_2)^2}{4\epsilon_1\epsilon_2}-\frac{\alpha^2}{\epsilon_1\epsilon_2}=\
\frac14(b+\frac1b)^2-\frac{\alpha^2}{g^2}$.
The quasiclassical limit corresponds to $b=g\rightarrow0$, which means $\epsilon_2=1,\epsilon_1\rightarrow0$. The SW limit corresponds to $c\ll\Delta$, so we should put $\frac bg\gg1$ and $g\ll1$, therefore in this limit both
$\epsilon_1\rightarrow0$, $\epsilon_2\rightarrow0$. In this limit our prepotential receives the $U(1)_\mathcal{R}$
symmetry, which was broken by some integer numbers in two-dimensional conformal theory.
It is still a nontrivial question about the limit of Painleve VI in such case.
We hope to return to this issue elsewhere.

\setcounter{equation}0
\section{Zamolodchikov's conformal blocks
\label{ss:AlZam}}

The AGT conjecture \cite{AGT} allows to apply the techniques of four-dimensional supersymmetric gauge
theories to answer to some complicated questions of two-dimensional conformal theory (see, e.g. \cite{BBFLT})
and vice versa. In the SW limit $\epsilon_1,\epsilon_2\rightarrow0$ one can identify the extended prepotentials to certain limit of the $c=1$ conformal blocks, and if the conformal dimensions are fixed
on the two-dimensional side, it just corresponds to vanishing of the masses of external multiplets.
The SW formulation, if $\Sigma_0$ is a sphere with punctures, leads to the set of differential equations for the conformal blocks in such limit, while the underlying geometry is the $g$-parametric family of genus $g$ curves.

It is interesting to compare this description for the $SU(2)$-quiver gauge theories with another well-known case, proposed by Al.~Zamolodchikov for the conformal blocks of $c=1$ Ashkin-Teller model \cite{ZamAT}, and described in very similar terms. The Zamolodchikov result for a
$2g+2$-point conformal block for the spin fields with external dimensions $\frac1{16}$
was given in terms of a \emph{a single} genus-$g$ curve, and required an extra charge-conservation
constrains for the dimensions in the internal vertices of the block.
Solving equation $2g+2=g+3$, one gets $g=1$, corresponding to the four-point conformal block and conformal $SU(2)$ supersymmetric QCD \rf{confqcd}, \rf{prepn4}, where these two constructions obviously coincide.
In general situation, there is a difference, whose origin comes from the vanishing of some (half-) tri-fundamental masses - in the
triple-vertices. For the Zamolodchikov conformal blocks this is just charge conservation in $c=1$ conformal theory, which is certainly absent
for generic $SU(2)$ quiver theory on the gauge theory side. Hence, in the SW approach it is equivalent to the extra relations on the condensates
for three gauge groups, interacting with the same tri-fundamental multiplet of matter. The first time, when such conservation law can be imposed is the case of sicilian quiver with the curve \rf{siquicu} and the tau-function \rf{trifundexp}, the Zamolodchikov constraint is singular from the point of view of four-dimensional physics (vanishing of one of the multiplet masses, which has been already integrated out to get the SW effective action), but the prepotential \rf{trifundexp} is regular in this limit, and
becomes just a quadratic function of the condensates, in accordance with \cite{ZamAT}.

Another reason to discuss this case, which is explicitly solvable even on the CFT side of
the correspondence, is that there exists also the isomonodromic-CFT correspondence \cite{SJM,Painleve},
with an exact solution for the $2g+2$-point \emph{isomonodromic} tau-function of the special type
\cite{Kitaev-Korotkin}, related to the Zamolodchikov conformal block. So the constrained case
of the sicilian quiver and other gauge theories with massless fundamental and partially massless tri-fundamental matter turns to be exactly-solvable in three different approaches. Note also, that this case on gauge-theory side is the simplest example of the S-duality class, where the standard methods
of instanton calculus are not applicable \cite{nepes}, so the correspondence between the four-dimensional and two-dimensional sides goes in fact even \emph{beyond} the standard formulation \cite{AGT} of the AGT-correspondence.

A generic Zamolodchikov case corresponds to a special case of the $n=g+3$ point conformal
block with $V=|\mathcal{V}|=\half n -2$ triple vertices $\{ \mathcal{V}_i\in \mathcal{V}\}$ or half-tri-fundamental multiplets ($n$ must be even in this case). At each such vertex $c=1$
conformal theory gives
one conservation condition, so that the genus drops to
\be
\label{redg}
\tilde{g} = g-V = n-3 - \left(\half n -2\right) = \half n -1
\ee
and for the total number of punctures we restore $n = 2\tilde{g} +2$~\footnote{From now on we will denote by $\tilde{g}$ the genus of hyperelliptic curve in the constrained case. Due to the conservation conditions the number of remaining independent gauge theory condensates $\tilde{g}$ will be always less in Zamolodchikov's case than the amount $n-3 = 2\tilde{g}-1$ of the UV couplings.}.
Another form
\be
\tilde g-1=\frac{g-1}2
\ee
of the same relation \rf{redg} means that the Euler characteristic $\chi(\tilde{\Sigma}) = \half\chi(\Sigma)$ decreases twice after the degeneration.

We are going now to present the direct proof, that in such limit the extended SW prepotential \rf{SW0}, \rf{SW1} becomes the quadratic form
\be
\label{Zamprep}
\left.\mathcal{F}(\mathbf{a},\mathbf{q})\right|_{\bigcup_\mathcal{V}\sum_{i\in \mathcal{V}_i} a_i=0}=i\pi\sum_{\alpha,\beta=1}^{\tilde{g}} a_\alpha\mathcal{T}_{\alpha\beta}(\mathbf{q}) a_\beta
\ee
with the period matrix $\mathcal{T} = \|\mathcal{T}_{\alpha\beta}\|$ of the hyperelliptic
curve $\tilde{\Sigma}$ of genus $\tilde{g}$, which does not depend on the condensates (the moduli space of this hyperelliptic curve is parameterized by original set of the UV couplings). This result has been obtained originally, using the language of free field on Riemann surface.

Consider now the massless $SU(2)$ quiver theory with the generating differential
\be
\label{mslsgen}
dS = xdz =  \sqrt{\alpha}
{\sqrt{\prod_{k=1}^{g-1}(z-v_k)}\ dz\over \sqrt{\prod_{j=1}^{g+3}(z-z_j)}} \ \stackreb{(z_1,\ldots,z_{g+3})\to(q_1,\ldots,q_g,1,\infty,0)}{\longrightarrow}\
\\
\rightarrow\ \sqrt{\alpha}
{\sqrt{\prod_{k=1}^{g-1}(z-v_k)}\ dz\over \sqrt{z(z-1)\prod_{j=1}^{g}(z-q_j)}}
\ee
on a hyperelliptic curve \rf{quicur} of genus $g$, with the total number of branch points (from both numerator and denominator) $\# B.P. = 2g+2$. Impose now
\be
{g-1\over 2} = \half n -2 = V
\ee
constraints to the coefficients $\{ v_j \}$, $j=1,\ldots,g-1$ in the numerator of \rf{mslsgen} in order
to get the total square, i.e.
\be
\label{dSred}
dS = xdz \rightarrow {\mathcal{Q}_{\tilde{g}-1}(z)dz\over y}
\ee
with some polynomial $\mathcal{Q}_{\tilde{g}-1}(z)$ of power $\tilde{g}-1$,
which can be considered as a holomorphic differential on the ``reduced'' hyperelliptic curve $\tilde{\Sigma}$:
\be
\label{Zamcur}
y^2 = \prod_{j=1}^{g+3}(z-z_j)\ \stackreb{(z_1,\ldots,z_{g+3})\to(q_1,\ldots,q_g,1,\infty,0)}{\longrightarrow}\
\\
\rightarrow y^2 = z(z-1)\prod_{k=1}^g(z-q_k) = z(z-1)\prod_{k=1}^{2\tilde{g}-1}(z-q_k)
\ee
already of genus \rf{redg}. The differential \rf{dSred} can be decomposed
\be
\label{Zamdiff}
dS = {\mathcal{Q}_{\tilde{g}-1}(z)dz\over y} = \sum_{\alpha=1}^{\tilde{g}} a_\alpha{R_\alpha(z)dz\over y} = \sum_{\alpha=1}^{\tilde{g}} a_\alpha d\omega_\alpha
\ee
into a linear combination of the normalized holomorphic differentials on \rf{Zamcur}, so that the system of linear equations
\be
\label{Qa}
\frac1{2\pi i}\oint_{A_\alpha}{\mathcal{Q}_{\tilde{g}-1}(z)dz\over y} = a_\alpha,\ \ \ \ \alpha=1,\ldots,\tilde{g}
\ee
can be solved for $\tilde{g}$ coefficients of the polynomial $\mathcal{Q}_{\tilde{g}-1}(z)$. Equivalently, the system of equations
\be
\label{normZ}
\frac1{2\pi i}\oint_{A_\alpha}{R_\beta(z)dz\over y} = \delta_{\alpha\beta},\ \ \ \ \alpha,\beta=1,\ldots,\tilde{g}
\ee
fixes all $\tilde{g}^2$ coefficients of the polynomials $\{ R_\alpha(z)\}$ of power $\tilde{g}-1$, defining the normalized holomorphic differentials
\be
\label{normhol}
d\omega_\alpha = {R_\alpha(z)dz\over y},\ \ \ \ \alpha=1,\ldots,\tilde{g}
\ee
in \rf{Zamdiff}. The solution to the dual period equations
\be
\label{solvSW0}
{\d\F\over\d a_\alpha} = \oint_{B_\alpha}{\mathcal{Q}_{\tilde{g}-1}(z)dz\over y} =2\pi i\sum_{\beta=1}^{\tilde{g}} a_\beta \oint_{B_\alpha}d\omega_\beta
=2\pi i\sum_{j=1}^{\tilde{g}}   {\cal T}_{\alpha\beta}(\mathbf{q})a_\beta
\ee
immediately gives rise to the formula \rf{Zamprep} with the period matrix of \rf{Zamcur}, up to an $\mathbf{a}$-independent constant.
Relations to the dependence of the reduced prepotential upon the ramification points \rf{SW1}, i.e.
\be
\label{ZamSW1}
\frac{\d\mathcal{F}}{\d q_i} = \res_{q_i}\frac{(dS)^2}{dz} =
\frac{\mathcal{Q}^2_{\tilde{g}-1}(q_i)}{q_i(q_i-1)\prod_{j\neq i}(q_i-q_j)} =
\\
= \sum_{\alpha,\beta=1}^{\tilde{g}}
a_\alpha a_\beta \frac{R_\alpha(q_i)R_\beta(q_i)}{q_i(q_i-1)\prod_{j\neq i}(q_i-q_j)} ,
\ \ \ \ i=1,\dots,2\tilde{g}-1
\ee
immediately comes from \rf{Zamdiff} and completely fixed \rf{Zamprep} up to a constant.
This exactly coincides with the Zamolodchikov equation \cite{ZamAT} for the leading contribution to the correlator $e^\mathcal{F}=\langle\sigma_0(z_1)...\sigma_0(z_n)\rangle$ of spin fields in the AT model, see Appendix~\ref{ap:zamolod}. Below we shall also use it to prove the nonlinear relations, arising from the SW theory to the
derivatives of the matrix elements of the period matrix of hyperelliptic curves.

Results of this section are in complete agreement with the above analysis of the weak-coupling expansions
for the quiver tau-functions. Already from the perturbative part of \rf{trifundexp} we
see, that in Zamolodchikov's limit for $SU(2)\times SU(2)\times SU(2)$ prepotential
the expression for period matrix of $\Sigma$ becomes singular, when $a_1\pm a_2\pm a_3=0$ (vanishing mass of one of the states from the
(half-) tri-fundamental multiplet).  It means that the curve $\Sigma$ indeed
degenerates to $\tilde{\Sigma}$, and it is easy to see, that all denominators in \rf{trifundexp} disappear in this limit and the prepotential turns into a quadratic expression in the remaining SW periods.

The Zamolodchikov case extends the example of the $SU(2)$ conformal theory with elliptic curve \rf{confqcd} to a subfamily of quiver gauge theories which are non-renormalized within the perturbation theory, i.e. have vanishing beta-functions~\footnote{More strictly, the perturbative calculations give rise only
to a \emph{finite} renormalization of the couplings.}, but the true IR couplings are renormalized by
the non-perturbative effects. We are going to show in next section, that equations \rf{ZamSW1} are immediately rewritten in the form of differential equations for effective couplings (the derivatives are
taken over the bare couplings, since there are no other parameters in the theory), which take the form
of the Rauch relations, and can be implicitly solved via the Thomae formulas \cite{Fay,Mumford} (see
also \cite{KniMor}).

\setcounter{equation}0
\section{Non-linear equations in quiver gauge theory
\label{ss:eqs}}

In sect.~\ref{ss:wcexp} we have already used the differential equations, coming from the relations on
quasiclassical tau-functions \cite{KriW,AMtau}, to get the weak-coupling expansions for the
supersymmetric gauge theories. Particular examples of such equations (see e.g. \rf{schw}), and the direct
relation of these equations to quasiclassical expansions of the conformal blocks (and therefore to
Painleve VI) show that they have indeed some deep geometric origin.
Below in this section, we are going to study the differential equations, arising from the SW approach to quiver gauge theories, in more general context.

In the constrained Zamolodchikov's case all equations for the prepotential can be rewritten as relations
to the period matrices of hyperelliptic curves. We are going to show, that all such relations for
the first derivatives are actually consequences of the Rauch formulas. They propose \textit{some parametrization} in the space of first derivatives, which can be studied in algebro-geometric terms.

Another natural thing is to expect the WDVV-like equations \cite{MMM} to be satisfied
by extended prepotentials of the quiver gauge theories. We prove indeed, that such equations are
satisfied by quiver tau-functions in the massless case as functions of the whole set
of variables: all SW periods and bare couplings. Amazingly enough, the proof, based
on the residue formula \rf{residue} and a simple counting argument \cite{AMwdvv}, is valid both in the unconstrained and Zamolodchikov's cases,
leading in the latter situation to the relations including the third derivatives of the period matrices.

\subsection{Relations for the period matrix
\label{ss:relations}}

Consider, first, the simplest example of our SW system - the Zamolodchikov case of sect.~\ref{ss:AlZam}, represented by hyperelliptic curve \rf{Zamcur}, parameterized by the couplings $\mathbf{q}$ only, with the holomorphic SW differential \rf{Zamdiff}.
The SW equations \rf{SW0} are trivially solved, but the formula
(\ref{SW1}) is still non-trivial (see \rf{solvSW0}, \rf{Zamprep} and \rf{ZamSW1}).

Comparing the coefficients of the quadratic forms at the both sides of \rf{ZamSW1}, one gets for the
first derivatives of the period matrix of hyperelliptic curve \rf{Zamcur}
\be
\label{rauch}
\frac{\pd\mathcal{T}_{\alpha\beta}({\bf q})}{\pd q_k}=\mathrm{res}_{q_k}\frac{d\omega_\alpha d\omega_\beta}{dz}
= \frac{R_\alpha(q_k)R_\beta(q_k)}{q_k(q_k-1)\prod_{l\neq k}(q_k-q_l)},\ \ \ \ k=1,\ldots,2\tilde{g}-1
\ee
in terms of the numerators for the holomorphic differentials in \rf{normhol}: it is exactly one of the well-known Rauch formulas \cite{Fay}. Their solution can be found via the Thomae formulas \cite{Fay,Mumford},
which can be written for the curve \rf{Zamcur} in the form
\be
\label{thomae}
q_k^2 = \pm {\theta[\eta_1](\mathcal{T})^4\,\theta[\eta_2](\mathcal{T})^4
\over\theta[\eta_3](\mathcal{T})^4\,\theta[\eta_4](\mathcal{T})^4},\ \ \
k=1,\ldots,2\tilde{g}-1
\ee
for the set of four theta-characteristics, chosen in the following way. Divide the branch points as
\be
\{ z_1,\ldots,z_n\} = \{0,1,\infty,k\}\sqcup S' \sqcup S''
\ee
where $S'\supset\{ q_{j'}\}$ and $S''\supset\{ q_{j''}\}$ are any two nonintersecting sets, containing each $\tilde{g}-1$ branch points with $j'\neq k$, $j''\neq k$ and $j'\neq j''$. Then
\be
\eta_1 = \{S'\oplus k\oplus\infty\}\sqcup\{S''\oplus 0\oplus 1\}
\\
\eta_2 = \{S'\oplus 0\oplus 1\}\sqcup\{S''\oplus k\oplus \infty\}
\\
\eta_3 = \{S'\oplus 0\oplus k\}\sqcup\{S''\oplus 1\oplus \infty\}
\\
\eta_4 = \{S'\oplus 1\oplus \infty\}\sqcup\{S''\oplus 0\oplus k\}
\ee
are possible choices of even theta-characteristics in \rf{thomae} in terms of partitions of the
branch points. The proof of this fact can be found, for example, in \cite{Fay,KniMor}.

As an example, consider the first nontrivial Zamolodchikov's case with $\tilde{g}=2$, i.e.
\be
\label{degtozam}
x^2=\Sum_{i=1}^3\frac{q_i(q_i-1)}{z(z-1)(z-q_i)}\frac{\pd\mathcal{F}}{\pd q_i} = \frac{\alpha(z-z_0)^2}{z(z-1)(z-q_1)(z-q_2)(z-q_3)}
\ee
Expression in the r.h.s. means, that there
should one relation for the $\mathbf{q}$-derivatives of prepotential. It can be obtained by calculating the discriminant and leads to the algebraic equation
\be
\xi_1^2+\xi_2^2+\xi_3^2-2\xi_1\xi_2-2\xi_2\xi_3-2\xi_1\xi_3=0
\label{xi}
\ee
for the variables
\be
\xi_1=(q_2-q_3)q_1(q_1-1)\frac{\pd\mathcal{F}}{\pd q_1},\ \ \
\xi_2=(q_3-q_1)q_2(q_2-1)\frac{\pd\mathcal{F}}{\pd q_2}\\
\xi_3=(q_1-q_2)q_3(q_3-1)\frac{\pd\mathcal{F}}{\pd q_3}
\ee
and becomes an identity after using the Rauch formulas, or just substituting
\be
\xi_1=\frac{q_2-q_3}{(q_1-q_2)(q_2-q_3)}\mathcal{Q}(q_1)^2,\ \ \
\xi_2=\frac{q_3-q_1}{(q_2-q_1)(q_2-q_3)}\mathcal{Q}(q_2)^2\\
\xi_3=\frac{q_1-q_2}{(q_3-q_1)(q_3-q_2)}\mathcal{Q}(q_3)^2
\ee
for any linear $\mathcal{Q}_{\tilde{g}-1} = \mathcal{Q}(z)$.
Equation \rf{degtozam} also expresses
\be
\alpha=\frac{\pd\mathcal{F}}{\pd q_1}q_1(q_1-1)+\frac{\pd\mathcal{F}}{\pd q_2}q_2(q_2-1)+
\frac{\pd\mathcal{F}}{\pd q_3}q_3(q_3-1) = K_\alpha(a_1,a_2)
\label{coef_a}
\ee
and
\be
z_0=-\frac12\frac{\frac{\pd\mathcal{F}}{\pd q_1}q_1(q_1-1)(q_2+q_3)+\frac{\pd\mathcal{F}}{\pd q_2}
q_2(q_2-1)(q_1+q_3)+\frac{\pd\mathcal{F}}{\pd q_3}q_3(q_3-1)(q_1+q_2)}{
\frac{\pd\mathcal{F}}{\pd q_1}q_1(q_1-1)+\frac{\pd\mathcal{F}}{\pd q_2}q_2(q_2-1)+
\frac{\pd\mathcal{F}}{\pd q_3}q_3(q_3-1)}=
\\
= -\frac12\frac{K_z(a_1,a_2)}{K_\alpha(a_1,a_2)}
\label{coef_B}
\ee
in terms of quadratic forms in the SW periods with the coefficients
\be
\label{quadform}
K_\alpha^{ij}=\Sum_{k=1}^3q_k(q_k-1)\frac{\pd\mathcal{T}_{ij}}{\pd q_k}=\Sum_{k=1}^3q_k(q_k-1)\frac{\pd^3\mathcal{F}}{\pd q_k\pd a_i\pd a_j}
\\
K_z^{ij}=(q_1+q_2+q_3)K_\alpha^{ij}-\Sum_{k=1}^3q_k^2(q_k-1)\frac{\pd\mathcal{T}_{ij}}{\pd q_k}
=
\\
=(q_1+q_2+q_3)K_\alpha^{ij}-\Sum_{k=1}^3q_k^2(q_k-1)\frac{\pd^3\mathcal{F}}{\pd q_k\pd a_i\pd a_j}
\ee
However, the SW differential \rf{Zamdiff} for the curve \rf{degtozam}
\be
dS=\frac{\sqrt\alpha (z-z_0)}{\sqrt{z(z-1)(z-q_1)(z-q_2)(z-q_3)}}=a_1d\omega_1+a_2d\omega_2
\ee
states, that $\sqrt\alpha$ and $z_0\sqrt\alpha$ should be the linear functions of $a_1$ and $a_2$,
which results in equations
\be
\label{dopeq}
\det K_\alpha=0,\;\;\;\;\;\tr K_\alpha K_z^{-1}=0
\ee
for the \rf{quadform}. These equations, if considering them as constraints to the derivatives
of the matrix elements of the period matrices $\d_k\mathcal{T}_{\alpha\beta}$ should be considered independently of \rf{xi} (see Appendix~\ref{ap:intersect}).
Generally, all such relations just follow from representation of
\be
\label{Zamcurx2}
x^2=\Sum_{i=1}^{2\tilde{g}-1}\frac{q_i(q_i-1)}{z(z-1)(z-q_i)}\frac{\pd\mathcal{F}}{\pd q_i}\
\stackreb{\rf{ZamSW1}}{=}\ {1\over z(z-1)\prod_{j=1}^{2\tilde{g}-1}(z-q_j)}\sum_{k=1}^{2\tilde{g}-1}
\mathcal{Q}_{\tilde{g}-1}^2(q_k)\prod_{i\neq k}{z-q_i\over q_k-q_i}
\ee
where the sum in the r.h.s. is just the Lagrange interpolation formula for the polynomial $\mathcal{Q}_{\tilde{g}-1}^2(z)$ with vanishing discriminant.

\subsection{WDVV equations from residue formula}

Now let us show that prepotential of the $SU(2)$ quiver gauge theories satisfies the WDVV equations
\cite{WDVV} as the function of full set of variables  $\mathcal{F}=\mathcal{F}({\bf a},{\bf q})$. We have seen already, that in the case of $SU(2)$ gauge quivers the residue formula descends to the base-curve $\Sigma_0$ of the SW curve $\Sigma$, and - adjusting to this case - we reformulate the statement of \cite{AMwdvv} in the
following way:

\begin{theorem}Suppose that we have the formula
\be
\label{reswdvv}
\mathcal{F}_{IJK}=\Sum_{f(z)=0}\mathrm{res}\,\frac{r_I(z)r_J(z)r_K(z)}{f(z)}R(z)dz
\ee
where $f(z)$ is non-degenerate polynomial, $\deg f$ equals to the number of indices, $R(z)$ and $r_I(z)$ are
holomorphic at zeroes of the polynomial. Then there is a relation \cite{MMM} for the matrices\par\noindent
$(\mathcal{F}_I)_{JK}=\mathcal{F}_{IJK}$
\be
\label{MMM}
\mathcal{F}_I\mathcal{F}^{-1}_J\mathcal{F}_K=\mathcal{F}_K\mathcal{F}^{-1}_J\mathcal{F}_I
\ee
which is called the WDVV equation.
\end{theorem}

\textbf{Proof:} Define an auxiliary algebra $H_S$ of the functions on $N$ zeroes of $f(z)=\prod_{i=N}^n(z-\lambda_i)$ (which
is obviously isomorphic to $\mathbb{C}^N$) by the relation \begin{equation}
(r_I\mathop{*}\limits_{S} r_J)(\lambda_i)=S(\lambda_i)r_I(\lambda_i)
r_J(\lambda_i)\label{product}\end{equation}
and the homomorphism $l_S:H_S\rightarrow\mathbb{C}$ by

\be
\label{hom}
l_S(r)=\Sum_{i=1}^N\frac{R(\lambda_i)}{f'(\lambda_i)S(\lambda_i)^2}r(\lambda_i)
\ee
Compute the residue \rf{reswdvv} directly, substituting $f(z)=\prod_{i=N}^n(z-\lambda_i)$
\be
\mathcal{F}_{IJK}=\Sum_{i=1}^N\frac{R(\lambda_i)}{f'(\lambda_i)}r_I(\lambda_i)r_J(\lambda_i)r_K(\lambda_i)
= l_S(r_I\mathop{*}_S r_J\mathop{*}_S r_K)
\ee
Algebra $H_S$ is commutative and associative, and we can define  it's structure
constants \begin{equation}r_I\mathop{*}\limits_S r_J=\Sum_K C^K_{IJ}r_K
\end{equation}
 or the operators of multiplication by $r_I$ as $(C_I)^K_J=C^K_{IJ}$. Due to commutativity and associativity one has for these matrices
\be
C_I\cdot C_J=C_J\cdot C_I
\ee
Define now the bilinear form \begin{equation}\eta_{IJ}=l_S(r_I\mathop{*}\limits_S r_J)
\end{equation}
 which is obviously connected with
$(\mathcal{F}_I)_{JK}=\mathcal{F}_{IJK}$ by
\be
\mathcal{F}_I=\eta\cdot C_I
\ee
Then
\be
\label{WDVVeta}
\mathcal{F}_I\eta^{-1}\mathcal{F}_K = \mathcal{F}_K\eta^{-1}\mathcal{F}_J
\ee
Using the possibility to choose arbitrarily the function $S(\lambda)$ one can adjust it to make $r_J=1$
for some fixed $J$ (we simply take $S(\lambda_i)=\frac1{r_I(\lambda)}$ and look at (\ref{product})), then
$\eta=\mathcal{F}_J$, and \rf{WDVVeta} turns into \rf{MMM}.
\hfill $\Box$

{\bf Remark}: Our algebra $H_S$ is isomorphic to the algebra of functions on $N$ points, so it obviously contains the unity operator. Namely, take the linear combination of the basis elements $e_\alpha=\Sum_I\alpha_I r_I$ and then look at the multiplication
by this element:
\be
(e_\alpha\mathop{*}_S r_J)(\lambda_i)=\Sum_I \alpha_I r_I(\lambda_i) S(\lambda_i)r_J(\lambda_i) =
r_J(\lambda_i)
\ee
To ensure the last equality, one has just to solve the system of $N$ linear equations :
\be
\frac1{S(\lambda_i)}=\Sum_{I=1}^N\alpha_I r_I(\lambda_i),\,\,\,\, \lambda_i=1,...,N
\ee
for $N$ variables $\alpha_I$, with the only requirement $\det_{Ii}\|r_I(\lambda_i)\|\neq 0$. The
corresponding
\be
\eta_\alpha = \Sum_I\alpha_I\mathcal{F}_I
\ee
is natural to consider as bilinear form, corresponding to the unity operator, but there is no claim that
it does not depend on the dynamical variables.

\subsection{Proof for the quiver gauge theory}

As it was noticed, there are two important cases of the quiver gauge theories: ordinary and constrained one.
All considerations will be very similar, so we introduce the following short-hand notation: $(z-v_i)^{(1|2)}$
which means that we should read $(z-v_i)^1$ in the ordinary case and $(z-v_i)^2$ in Zamolodchikov case.

Apply now residue formula (\ref{residue}) to the particular case of
the massless $SU(2)$ quiver gauge theory (\ref{quicur}), rewriting it, first, for vanishing $\{\Delta_i\}$
in the form
\be
x^2=\frac{\alpha\prod_{s=1}^{(g-1|\frac{g-1}2)}(z- v_s(q_i,a_i))^{(1|2)}}{z(z-1)\prod_{k=1}^{g}(z-q_i)}
\ee
Express the corresponding $\mathbf{q}$- and $\mathbf{a}$- derivatives of $dS = xdz$ as:
\be
\label{alldiff}
d\Omega_k=\frac{\pd }{\pd q_k}xdz=\left(\frac12\frac1{z-q_k}+\Sum_{s=1}^{(g-1|\frac{g-1}2)}\frac{c_s}{z- v_s}+\frac12
\frac{\pd\log\alpha}{\pd q_k}\right)x dz=\mathfrak{R}_k(z) xdz
\\
d\omega_i=\frac{\pd }{\pd a_i}xdz=\left(\Sum_{s=1}^{(g-1|\frac{g-1}2)}\frac{b_s}{z- v_s}+
\frac12\frac{\pd\log\alpha}{\pd a_i} \right)xdz=\mathfrak{r}_i(z)xdz
\ee
where $i=1,\ldots,(g|\frac{g+1}2)$, $k=1,\ldots,g$, and unify now all variables into a single set $\{ X_I\}=\{ a_i\}\cup\{ q_k\}$, $\{ d\varpi_I\}=\{ d\omega_i\}\cup\{ d\Omega_k\}$, $\{ r_I\}=\{ \mathfrak{r}_i\}\cup\{\frak R_k\}$, with $I=1,\ldots,(2g|\frac{3g+1}2)$. The residue formula \rf{residue} gives rise to
\be
\frac{\pd^3\mathcal{F}}{\pd X_I\pd X_J\pd X_K}=2\Sum_{dx=0}\res\frac{d\varpi_I d\varpi_J d\varpi_K}
{dx dz}=
- 2\Sum_{dz=0}\res\,{\frac{\pd\log x}{\pd X_I}\frac{\pd\log x}{\pd X_J}\frac{\pd\log x}{\pd X_K}
\over \frac1{x^2}\frac{d\log x}{dz}}dz
\ee
where in the denominator of the r.h.s. we get explicitly
\be \frac1{x^2}\frac{d\log x}{dz}=\frac12\left(-\frac1z-\frac1{z-1}-\Sum_{i=1}^{g}\frac1{z-q_i}+\Sum_{s=1}^{(g-1|\frac{g-1}2)}
\frac{(1|2)}{z- v_s}\right)
\frac{z(z-1)\prod_{i=1}^{g}(z-q_i)}{\alpha\prod_{s=1}^{(g-1|\frac{g-1}2)}(z- v_s)^{(1|2)}}=
\\
= -\frac{2f_{(2g|\frac{3g+1}2)}(z)}{\prod_{s=1}^{(g-1|\frac{g-1}2)}(z- v_s)^{(2|3)}}
\ee
where $f_{(2g|\frac{3g+1}2)}(z)$ is a polynomial of degree $(2g|\frac{3g+1}2)$. Therefore
\be
\frac{\pd^3\mathcal{F}}{\pd X_I\pd X_J\pd X_K}=\Sum_{f_{(\bullet)}(z)=0}\res\,\frac{r_I(z)
r_J(z)r_K(z)}{f_{(2g|\frac{3g+1}2)}(z)}\prod_{s=1}^{(g-1|\frac{g-1}2)}(z- v_s)^{(2|3)}dz
\label{res1}
\ee
and since $r_I(z)$ are all holomorphic at zeroes of $f_{(2g|\frac{3g+1}2)}(z)$, and the total number of variables
is $(2g|\frac{3g+1}2)$, one can immediately apply \textbf{Theorem 4}. It means, that we have proven, that the WDVV equations hold for the SW tau-function, as function of all periods and couplings, for both
constrained and unconstrained case on equal footing.

\setcounter{equation}0
\section{Conclusion
\label{ss:concl}}

In this paper we have studied in detail the properties of the $SU(2)$ quiver gauge theories, along the lines proposed in \cite{AMtau}. We have derived and proved the residue formula for the third derivatives,
and used it for some further applications.

We have shown, that the residue formula provides an effective way for the computation of the weak-coupling expansion of the quiver gauge theory prepotentials. These residue formulas can be used as a differential equation, which can
be solved recursively, and this is equivalent to the expansion of the SW periods - the integrals of motion for these differential equations. Another application of the residue formula is that it leads immediately to the WDVV equations for the extended prepotential, once the number of critical points is equal to the number of deformations. We have checked above, that this condition holds both in the case of the full quiver theory, and its restricted or Zamolodchikov's case.

The Zamolodchikov case has attracted our special attention. We have completely described it in the SW approach for the quiver theory, showing that it arises after constraints, corresponding to arising
of a massless state of a tri-fundamental matter. The prepotential then turn to be quadratic expressions in SW periods, forming a new class of conformal gauge theories, where the bare UV couplings are
corrected only non-perturbatively. This renormalization can be described in terms of the Thomae formulas
for the branching points of hyperelliptic curves, which generalize naturally the Zamolodchikov renormalization in the single $SU(2)$ conformal theory.

For the
higher rank gauge groups the situation seems to be far more complicated, but it looks like it can be studied by the methods, proposed in this paper.
The extension to the higher-rank gauge theories can be possible started with extension of the Zamolodchikov's case, whose SW formulation is one of the main results of the present paper. Complete analysis of the higher rank case requires also the study of
the higher Teichm\"uller spaces and
corresponding deformations of the UV gauge theory, but the higher rank analogs of the Zamolodchikov case should be understandable in the SW terms, since on the CFT side it is described in terms of a systems of several scalar fields on Riemann surfaces. We plan to return to this problem elsewhere.

\section*{Appendix}
\appendix

\setcounter{equation}0
\section{Conformal block in the Ashkin-Teller model
\label{ap:zamolod}}

Here we present the simplified derivation of the exact conformal block in $c=1$ AT model in terms convenient for the correspondence with the SW theory.
The starting point in \cite{ZamAT} is the operator algebra of the AT model which contains:
\begin{itemize}
\item The $U(1)$-current $I(z)$, the Sugawara stress-energy tensor is $T(z)= :I(z)^2:$;
\item The Virasoro primary spin field $\sigma_0(z)$ and its first descendant in the current module $\sigma_1(z)$, which are $\mathbb{Z}_2$ twist-fields in terms of $I(z)$ (do not have the $U(1)$ charge).
\end{itemize}

Consider the conformal or ``current blocks'' of the spin fields $\langle\sigma_0(z_1)\ldots\sigma_0(z_n)\rangle$, where the charges in the intermediate channel
is fixed by
\be
\frac1{2\pi i}\Oint_{A_\alpha} I(z)dz=a_\alpha,\ \ \ \ \alpha = 1,\ldots,\tilde{g}={n\over 2}-1
\label{charges}
\ee
where the $A$-cycles encircle each two spin fields, and are interpreted as canonical on the
hyperelliptic curve
\be
\label{hyZam}
y^2=\prod_{i=1}^{2\tilde{g}+2}(z-z_i)
\ee
introduced to make the correlator of spin-fields to be a single-valued function on this two-fold cover of the initial sphere $\Sigma_0$. The operator product expansions \cite{ZamAT}
\be
I(z)\sigma_0(0)=\half z^{-\frac12}\sigma_1(0)+\ldots
\\
I(z)\sigma_1(0)=\half z^{-\frac32}\sigma_0(0)+2z^{-\frac12}\pd\sigma_0(0)+\ldots
\label{fusion}
\ee
predict for the ratios of the correlation functions
\be
F_0\left(z|\{z_i\}\right)=\frac{\langle I(z)\sigma_0(z_1)\ldots\sigma_0(z_{2\tilde{g}+2})\rangle}
{\langle\sigma_0(z_1)\ldots\sigma_0(z_{2\tilde{g}+2})\rangle} =\frac{{\cal Q}_{g-1}(z)}y
\label{F0}
\ee
where the r.h.s. is written from the (\ref{fusion}) and contains a $g$-parametric polynomial, totally fixed by the period integrals \rf{Qa}, coming now from
\rf{charges}. For the ratio of slightly different correlation functions it follows from \rf{fusion} and analytic properties
\be
F_1\left(z|\{z_i\}\right)=\frac{\langle I(z)\sigma_1(z_1)\sigma_0(z_2)\ldots\sigma_0(z_{2\tilde{g}+2})\rangle}{\langle \sigma_1(z_1)\sigma_0(z_2)\ldots\sigma_0(z_{2\tilde{g}+2})\rangle}=\frac{{\cal Q}_{g-1}(z)}y+
\frac{d\Omega_1}{dz}
\label{F1}
\ee
where
\be d\Omega_1=\frac{C_1dz}{y}\left({1\over z-z_1}+O_{g-1}(z)\right),\ \ \ \ \Oint_{A_i}d\Omega_1=0
\ee
is the normalized Abelian differential on \rf{hyZam}. The operator product expansion
(\ref{fusion}) inserted into \rf{F0} at $z\to z_1$ leads to relation
\be
\frac{{\cal Q}_{g-1}(z_1)}{\sqrt{\prod_{j\neq1}(z_1-z_j)}}
\langle\sigma_0(z_1)\sigma_0(z_2)\ldots\sigma_0(z_{2\tilde{g}+2})\rangle = \half\langle\sigma_1(z_1)\sigma_0(z_2)\ldots\sigma_0(z_{2\tilde{g}+2})\rangle
\label{eq1}
\ee
while inserted into (\ref{F1}) gives
\be
\frac{\langle\sigma_0(z_1)\sigma_0(z_2)\ldots\sigma_0(z_{2\tilde{g}+2})\rangle}
 {\langle\sigma_1(z_1)\sigma_0(z_2)\ldots\sigma_0(z_{2\tilde{g}+2})\rangle} = \frac{2C_1}{\sqrt{\prod_{j\neq1}(z_1-z_j)}}
\ee
i.e. $4C_1{\cal Q}_{g-1}(z_1) = \prod_{j\neq1}(z_1-z_j)$, and
\be
2\frac{\d_{z_1}\langle\sigma_0(z_1)\sigma_0(z_2)\ldots\sigma_0(z_{2\tilde{g}+2})\rangle}
 {\langle\sigma_1(z_1)\sigma_0(z_2)\ldots\sigma_0(z_{2\tilde{g}+2})\rangle} =
\frac{{\cal Q}_{g-1}(z_1)+C_1O_{g-1}(z_1)}{\sqrt{\prod_{j\neq1}(z_1-z_j)}} - {C_1\over 2}\sum_{j\neq 1}{1\over (z_1-z_j)^{3/2}}
\ee
These two formulas together result in
\be
\label{Zameq}
\left(2\pd_{z_1}-\frac{2{\cal Q}_{g-1}^2(z_1)}{\prod_{j\neq 1}(z_1-z_j)}+\frac14\sum_{j\neq 1}\frac1{z_1-z_j}+\frac12O_{g-1}(z_1)\right)
\langle\sigma_0(z_1)\sigma_0(z_2)\ldots\sigma_0(z_{2\tilde{g}+2})\rangle=0
\ee
Substituting the anzatz $\langle\sigma_0(z_1)\sigma_0(z_2)\ldots\sigma_0(z_{2\tilde{g}+2})\rangle=e^\mathcal{F({\bf a},{\bf z})}\cdot G({\bf z})$,
where only the function in the exponent is $\mathbf{a}$-dependent, one can extract from \rf{Zameq}
the equality
\be
\frac{\pd\mathcal{F}}{\pd z_i}=\frac{{\cal Q}_{g-1}^2(z_i)}{\prod_{j\neq i}(z_i-z_j)} 
\ee
which coincides exactly with \rf{ZamSW1}. It is also clear, that formulas \rf{charges}, \rf{Qa} turn into the first half of the periods of the SW differential. Hence, the exponential $\mathbf{a}$-dependent
contribution to the solution \rf{Zameq} can be obtained using the techniques presented in the main text of the paper, which leads immediately to the answer \rf{Zamprep}, where the quadratic form is already identified with the period matrix of \rf{hyZam}, while it has been established only with some additional argumentation in the original paper \cite{ZamAT}.

\setcounter{equation}0
\section{More on derivatives of the period matrices
\label{ap:intersect}}

Here we present some analysis of the equations \rf{xi} and \rf{dopeq}, following from the Rauch
formulas \rf{rauch}. Forgetting about the normalization conditions \rf{normZ}, one can consider
equations \rf{rauch} as parametrization of some submanifold in the space $\mathbb{A}^{\frac12\tilde g(\tilde g+1)(2\tilde g-1)}$ of
the derivatives
of our period matrix, where one can express the coordinates
\be
T^k_{\alpha\beta}=\prod_{l\neq k}(q_k-q_l)\frac{\pd\mathcal{T}_{\alpha\beta}}{\pd q_k}
\ee
in terms of the ${\tilde g}^2$ coefficients $\{R_{\alpha\beta}\}$ of the polynomials $R_\alpha(z)$. From
this point of view formulas \rf{rauch} define
the map $\mathbb{A}^{\tilde g^2}\ \mapsto \mathbb{A}^{\frac12\tilde g(\tilde g+1)(2\tilde g-1)}$
by quadratic functions
\be
T^k_{\alpha\beta}=R_\alpha(q_k)R_\beta(q_k)
\label{map}
\ee
so it can
be considered as the map $\mathbb{P}^{\tilde g^2-1}\mapsto\mathbb{P}^{\frac12\tilde g(\tilde g+1)(2\tilde g-1)-1}$.

The question, which allows to understand better the origin of equations \rf{xi}, \rf{dopeq} is how to describe the image of this map. In the $\tilde g=2$ case, where we get
$\mathbb{P}^{\,3}\mapsto\mathbb{P}^{\,8}$, the codimension is five and one needs at least five equations in $\mathbb{P}^{\,8}$. Notice, that we have already five independent equations in \rf{xi}, but
\rf{dopeq} give two extra.

Parameterizing our $\tilde g=2$ polynomials explicitly
\be
R_1(z)=az+b,\,\,\,\,\,R_2(z)=cz+d
\ee
where $(a:b:c:d)$ are the homogeneous coordinates on $\mathbb{P}^3$, one can write
\be
T^i_{11}=(aq_i+b)^2,\,\,\,\,T^i_{12}=(aq_i+b)(cq_i+d),\,\,\,\,T^i_{22}=(cq_i+d)^2,\ \ \ \
i=1,2,3
\ee
Since all equations are quadratic in $a,b,c,d$ the
intersection with the general codimension-3 plane
contains $2^3=8$ points, so the degree of the image is 8.

Suppose now, that the image is a total intersection in $\mathbb{P}^{\,8}$, then it should be determined by five polynomials $P_{d_1},\ldots,P_{d_5}$, such
that $d_1\cdot\ldots\cdot d_5=8$. It is possible only if at least two powers are $d_i=1$, therefore the image lies in a hyperplane.
However, this turns to be impossible, since in such case one gets a linear equation
\be
\Sum_{\alpha,\beta,k}C_{\alpha\beta}^kT^k_{\alpha\beta}=0
\ee
which is immediately rewritten as
\be
C_{aa}a^2+C_{ab}ab+\ldots+C_{dd}d^2=0
\ee
true $\forall(a,b,c,d)$, but this is impossible if $C_{IJ}\neq 0$. Hence, we come to a contradiction, and our surface does not lie in the hyperplane.
In particular, it means, that the equations \rf{dopeq} are not the consequence of \rf{xi}, and should be considered independently.

\section*{Acknowledgements}

We are grateful to I.~Krichever, A.~Morozov and T.~Shabalin for useful discussions, and to N.~Nekrasov and V.~Pestun for their comments of related topics and critical remarks. The preliminary results
have been reported at the Russian-Japanese JSPS/RFBR workshop on Integrability and
Gauge/String Duality in Moscow
and at the workshop on Integrability and Gauge Theory, held within the program ``Cohomology in Mathematics and Physics'' of the Euler Institute in St.-Petersburg in September-October 2013.

The work  was carried out within the research grant 13-05-0006
of the NRU HSE Academic Program support in 2013, and joint Ukrainian-Russian (NAN-RFBR) project 01-01-14 (P.G.) and 14-01-90404 (A.M.). The work of P.G. has been also supported by the grant of Laboratory of the Algebraic Geometry (HSE). The work of A.M. has been also supported by RFBR grant 14-01-00547, by joint RFBR/JSPS project 12-02-92108, by the Program of Support of Scientific Schools (NSh-1500.2014.2), and by the Russian Ministry of Education.

\end{document}